\documentclass[11pt]{article}
\pdfoutput=1
\usepackage{jheppubmod}
\usepackage{color}
\usepackage[showonlyrefs]{mathtools}
\usepackage{psfrag}
\usepackage{array}
\usepackage{amssymb}
\usepackage{graphicx}
\usepackage{tikz}
\usetikzlibrary{trees}
\usetikzlibrary{snakes}
\usetikzlibrary{arrows}
\usepackage{amsmath}
\usepackage{empheq}
\usepackage{slashed}
\usepackage{caption}
\usepackage[labelsep=quad]{subcaption}
\usepackage{epstopdf}
	
\usepackage{epsfig}
\usepackage[punctsep]{collref}
\newcommand{\be}{\begin{equation}}
\newcommand{\ee}{\end{equation}}
\collectsep[]{;}	
\usepackage{cite}
 
\allowdisplaybreaks

\def\FF{{\cal F}}

\def\HH{{\cal H}}

\def\KK{{\cal K}}
\def\LL{{\cal L}}
\def\MM{{\cal M}}
\def\NN{{\cal N}}
\def\OO{{\cal O}}

\def\SS{{\cal S}}

\def\Tr{{\rm {Tr}}}

\def\beq{\begin{equation}}
\def\eeq{\end{equation}}
\newcommand{\bea}{\begin{eqnarray}}
\newcommand{\eea}{\end{eqnarray}}

\newcommand{\RT}{Ryu-Takayanagi }
\newcommand{\rad}{\sigma}
\newcommand{\nn}{\nonumber}
\newcommand{\surf}{\gamma}
\newcommand{\surfRT}{\gamma_{RT}}
\newcommand{\radz}{\rad_0}
\newcommand{\mapz}{r_{\rad_0}}

\newcommand{\eLL}{\ell}
\newcommand{\separatrix}{separatrix\ } 
\newcommand{\UNO}{{\bf A}}
\newcommand{\DUE}{{\bf B}}
\newcommand{\Func}{F}
\newcommand{\yc}{{\hat \sigma}}
\newcommand{\X}{X}
\newcommand{\bz}{\bar{z}}
\newcommand{\etair}{\eta_{IR}}
\newcommand{\yh}{s}

\title{Large-$N$ transitions of the connectivity index}

\author{Francesco Aprile,}
\author{Vasilis Niarchos}
\affiliation{Crete Center for Theoretical Physics and\\ 
Crete Center for Quantum Complexity and Nanotechnology,\\
Department of Physics, University of Crete, 71303, Greece}
\emailAdd{aprile@physics.uoc.gr}
\emailAdd{niarchos@physics.uoc.gr}

\preprint{CCQCN-2014-47
\\ \hspace*{\fill}
CCTP-2014-22\\\hspace*{\fill}
}

\date{}

\abstract{The connectivity index, defined as the number of decoupled components of a separable
quantum system, can change under deformations of the Hamiltonian or during the dynamical change of the 
system under renormalization group flow. Such changes signal a rearrangement of correlations 
of different degrees of freedom across spacetime and field theory space. In this paper we quantify 
such processes by studying the behavior of entanglement entropy in a specific example: 
the RG flow in the Coulomb branch of large-$N$ superconformal field theories. 
We find evidence that the transition from the non-separable phase of the Higgsed gauge theory in the UV
to the separable phase of deformed decoupled CFTs in the IR exhibits sharp features in the middle of the 
RG flow in the large-$N$ limit. The entanglement entropy on a sphere with radius $\ell$ exhibits the formation 
of a separatrix on the co-dimension-two Ryu-Takayanagi surface in multi-centered brane geometries above 
a critical value of $\ell$. We discuss how other measures of entanglement and separability based on the 
relative quantum entropy and quantum mutual information might detect such transitions between non-separable 
and separable phases and how they would help describe some of the key properties of the IR physics of such flows. 
}


\begin{document}
\maketitle

\section{Introduction}
\label{intro}

In any quantum system we can arbitrarily partition the total Hilbert space $\HH$ into two subspaces 
$\HH_1$ and $\HH_2$. For a given configuration we can ask to what extent states in $\HH_1$ are 
entangled with states in $\HH_2$, or how strongly observables computed in $\HH_1$ are 
correlated with observables in $\HH_2$. This is an interesting question that can reveal useful information 
about the state of the system and its dynamical properties.

In a well studied example we take a system defined in $p$ spatial dimensions and separate the degrees of freedom
inside a spatial region $A$ from the degrees of freedom in the complement $A^c$. A natural measure of the 
entanglement of the two sets of degrees of freedom (in Hilbert spaces $\HH_A$ and $\HH_{A^c}$) is the 
entanglement entropy defined as the von-Neumann entropy of the reduced density matrix $\rho_A$
\beq
\label{introaa}
S = -\Tr_{\HH_A} \left[ \rho_A \log \rho_A \right]
~.
\eeq
$\rho_A$ is obtained by tracing the total density matrix $\rho$ over the states of the outside Hilbert space 
$\HH_{A^c}$
\beq
\label{introab}
\rho_A = \Tr_{\HH_{A^c}} \rho
~.
\eeq
$S$ is an interesting quantity that has played a central role in many recent developments. For example, 
when applied to $(p+1)$-dimensional relativistic conformal field theories its dependence on the characteristic
size of the region $A$ holds information about basic constants of the theory, $e.g.$ the central charge 
$c$ in $(1+1)$ dimensions \cite{Holzhey:1994we} (see e.g. \cite{Calabrese:2009qy} for a review), 
or the $F$-function in $(2+1)$ dimensions \cite{Myers:2010tj} etc.

Another possibility is to partition the system in field theory space, namely split the degrees of freedom at 
each point of spacetime into two subsets. This type of partitioning arises naturally, for example, when we 
have two distinct quantum systems with Hamiltonians $H_1$ and $H_2$ interacting weakly via an interaction 
Hamiltonian $H_{int}$, but it can also be considered more generally without reference to a specific type of 
dynamics.

The first question we want to ask in this paper is the following. Given an arbitrary split of the degrees of 
freedom of a quantum system, e.g.\ a quantum field theory, in spacetime and/or in field theory space, can we 
define a meaningful measure of the entanglement or strength of correlation between the subsystems.
Several well known measures from quantum information theory that quantify the notion of 
separability, e.g.\ measures based on the relative quantum entropy, turn out to be very well suited for 
this purpose. We will review the relevant concepts, and give specific definitions, in section \ref{entangle}.

The second question we want to raise concerns the behavior of such measures under deformations of
the theory, or under the dynamical change of the parameters of the system under the renormalization 
group (RG) flow. 

For example, there are many cases where by tuning the parameters of a theory, or by looking at the system at 
different energies, the interaction coupling in $H_{int}$ between two subsystems becomes weak or even turns off. 
In the latter case the subsystem Hamiltonians $H_1$ and $H_2$ decouple completely. Any observable computed 
in this product theory (e.g.\ an arbitrary correlation function) factorizes in a (sum of) products of observables of 
theory 1 and  theory 2. It is useful to introduce a {\it connectivity index}\footnote{An analogous concept was considered 
previously in \cite{Aharony:2006hz}. In that work a rough definition of the connectivity index was presented in terms 
of the independent gauge groups of a gauge theory (see also \cite{Kiritsis:2006hy} for very closely related work). 
Here we try to define the connectivity index in a more general (not necessarily equivalent) manner.} 
that quantifies how many independent parts a quantum system possesses. 
Along the lines of factorizability, one might define the connectivity index to be $n$, if the arbitrary correlation 
function factorizes in a (sum of) products of correlation functions of $n$ independent subsystems.
Employing the concept of separability from quantum information theory, one could 
alternatively define the connectivity index as the number of separable components
of the density matrix of the system (for a definition of separable density matrices see section \ref{entangle}).
Yet another natural definition is the following. Notice that a theory with $n$ decoupled 
components will have in general $n$ independently conserved energy-momentum tensors. 
This suggests defining the connectivity index as the number of conserved energy-momentum 
tensors. In the examples that we consider the above definitions appear to be equivalent, but we do 
not have a clean proof. Their relation is discussed further in section \ref{moreIR}.

With any of the above definitions the connectivity index can decrease when $H_{int}$ turns on, or increase when $H_{int}$ turns off. 
The measures of entanglement mentioned above will behave accordingly. It is possible, however, to
encounter more subtle situations where many of the effects mediated by $H_{int}$ are 
suppressed until a finite value of the interaction coupling. We will argue that RG flows in the Coulomb 
branch of large-$N$ superconformal field theories (SCFTs) provide interesting examples of this type. 

For definiteness, let us consider the Coulomb branch of the four-dimensional $SU(N)$ $\NN=4$ 
super-Yang-Mills (SYM) theory. In the ultraviolet (UV) we have an $SU(N)$ gauge theory with 
the apparent connectivity index 1. Turning on the vacuum expectation values of the adjoint scalars we move away 
from the origin of the Coulomb branch, the gauge group is Higgsed, say to $SU(N_1) \times SU(N_2)\times U(1)$, 
and there is an RG flow to the infrared (IR) where an $SU(N_1)$ gauge theory decouples from an $SU(N_2)$ gauge theory. 
In the far IR the connectivity index counts 2 decoupled components with order $N^2$
degrees of freedom and another component associated with the decoupled degrees of freedom
of the $U(1)$ part of the theory. At low energies the leading order direct interaction between the two $SU(N_i)$ IR CFTs is 
mediated by an irrelevant double-trace dimension 8 operator \cite{Intriligator:1999ai,Costa:1999sk}
(see section \ref{multicenterIR} for specific expressions). Being irrelevant this operator turns off at the extreme IR.
As we explain in section \ref{fragment} the $U(1)$ part also mediates interactions and plays an interesting role in the low energy dynamics.

The interest in the large-$N$ limit stems from the following observation. 
If we could isolate the dynamics of the $SU(N_i)$ IR CFTs from the dynamics of
the $U(1)$ part, we would be able to argue at leading order in the $1/N$ expansion that
the multi-trace operators that mediate interaction between the two IR CFTs do 
not contribute to the anomalous dimension of any combination of their energy-momentum 
tensors and despite the deformation both energy-momentum tensors remain independently conserved.
That would be evidence that the system remains in a separable state in a {\it vicinity} of the IR fixed
point at leading order in $1/N$. In the actual RG flow, however, one cannot isolate the dynamics of the $U(1)$ part. 
Since the latter mediates interactions that allow energy to flow from the $SU(N_1)$ to the $SU(N_2)$
IR CFTs the system is expected to be in a non-separable state with connectivity index 1 
infinitesimally away from the extreme IR. It is interesting to find a quantitative measure
that expresses how strongly the IR separability is broken by such ($U(1)$-mediated) 
interactions and to explore how one connects this type of infrared physics to the UV physics 
of a strongly non-separable Higgsed gauge theory in the UV.

One observable that we consider in the main text, in order to examine these questions, is the entanglement 
entropy \eqref{introaa} for a spherical geometry $A$ with radius $\ell$. As $\ell$ changes from 0 to $+\infty$ 
we probe physics from the UV to the IR. In the large-$N$ limit we can evaluate the entanglement entropy 
with the generalized Ryu-Takayanagi prescription \cite{Ryu:2006bv,Nishioka:2006gr,Klebanov:2007ws}
by determining a minimal co-dimension-2 hypersurface in the multi-throat geometry of separated 
stacks of branes. We perform this analysis quite generally for the 4d $\NN=4$ SYM theory on D3 branes, 
the 3d ABJM theory on M2 branes and for the 2d CFT on D1-D5 branes. In all cases we find that the 
Ryu-Takayanagi minimal hypersurface exhibits a separatrix at a radius $\ell_{c}$ where it shows
signs of critical behavior. This is evidence of an interesting sharp feature that occurs in the middle of the RG flow.

In section \ref{entangle} we define other measures of entanglement based on the concept
of relative quantum entropy. Currently, we do not know how to compute these measures
holographically from gravity in the large-$N$ limit, but we discuss possible behaviors in sections \ref{moreIR} and \ref{discussion}.
Eventually, one would like to determine how these measures capture
the quantum field theory dynamics that is summarized in section \ref{fragment}.

The main computational results of the paper are presented in sections \ref{multicenter}-\ref{HEE2ir}.
Section \ref{HEEgeneric} contains a description of the qualitative features of the \RT surface
in multi-centered geometries. The reader can consult this section for a quick overview of the 
results that arise by studying the holographic entanglement entropy in this context. Concrete 
quantitative results based on the analysis of the equations of the Ryu-Takayanagi minimal surface are 
presented in sections \ref{HEE2uv}, \ref{HEE2ir}. For instance, in section \ref{HEE2uv} we notice that 
the UV expansion of the holographic entanglement entropy does not receive contributions from the 
lowest order harmonics. This is a gravity prediction for a corresponding statement about entanglement 
entropy in the large-$N$ superconformal field theories that we consider. 

Interesting aspects of our story and other open issues are summarized and further discussed 
in the concluding section \ref{discussion}. Useful technical details are relegated to appendix \ref{APPEQ}.

\section{Separability, relative quantum entropy and other useful concepts}
\label{entangle}

Assume that we have a $(p+1)$-dimensional quantum system with Hilbert space $\HH$ and we 
partition $\HH$ both in spacetime and field theory space. In spacetime we separate states 
supported inside a spatial region $A$ from states in the complement $A^c$. In field theory 
space, we separate (at each point of spacetime) degrees of freedom of a subsystem 1 from 
degrees of freedom of a subsystem 2. Then, the reduced density matrix $\rho_A$ \eqref{introab} 
is a matrix that lives in the product Hilbert space $\HH_{A,1} \otimes \HH_{A,2}$. We are interested 
in a measure that quantifies the entanglement of the states of the two subsystems 1 and 2. We 
will focus on the properties of the density matrix $\rho_A$ keeping the additional dependence 
on the size of the region $A$ as a useful way to keep track of the entanglement across different 
length (or energy) scales.

A standard definition in quantum information theory (see \cite{wilde} for a review) postulates that 
the state represented by $\rho_A$ is {\it separable} if it can be written as a sum of product states 
in the form
\beq
\label{enteaa}
\rho_A = \sum_k p_k\, \rho_{A,1}^{(k)} \otimes \rho_{A,2}^{(k)}
~,
\eeq
with $p_k \geq 0$, $\sum_k p_k =1$. If not, $\rho_A$ is called entangled. In the special case with 
a single propability coefficient $p_k$ non-zero, i.e.\ when
\beq
\label{entab}
\rho_A = \rho_{A,1} \otimes \rho_{A,2}
\eeq
the state is called {\it simply separable}. This is the case mentioned in the introduction where
no correlations between subsystems 1 and 2 exist.

Testing for separability is in general a very hard problem. However, it is possible to formulate a measure
that quantifies how far from separability a quantum system is by using the concept of relative quantum entropy.
For any two density matrices $\rho,\sigma$ the relative quantum entropy of $\rho$ with respect to $\sigma$
is defined as 
\beq
\label{entac}
S(\rho \, ||\, \sigma)= \Tr \left[ \rho \log \rho \right] - \Tr \left[ \rho \log \sigma \right]
~.
\eeq
One can prove the Klein inequality (see e.g.\ \cite{wilde}), which states that $S(\rho \, ||\, \sigma)$ is a
positive-definite quantity that vanishes only when $\rho=\sigma$, i.e.\ when the states $\rho$ and
$\sigma$ are indistinguishable. On the other extreme, the relative quantum entropy 
$S(\rho \, ||\, \sigma)$ is infinite when the two states are perfectly distinguishable. This fact played
a useful role in the recent work \cite{Hartnoll:2014ppa}.

One can use the relative quantum entropy to define a measure of how far a system is from separability.
The usual approach defines the following quantity
\beq
\label{entad}
D_{\rm REE}(\rho) = \min_{\sigma\, = \, separable} S(\rho\, || \, \sigma)
~,
\eeq
which is called relative entropy of entanglement. The minimum is obtained by sampling over the whole space
of separable states. $D_{\rm REE}(\rho)$ is zero if and only if $\rho$ is a separable state.

Since we are interested in simply separable states we can modify this definition in an obvious way 
by taking the minimum over the simply separable states. In what follows, however, we consider instead 
a related quantity that we define as follows. Concentrating on the specific context of our density matrix 
$\rho_A$, and a partitioning into two complementary subsystems 1 and 2, we consider 
the relative quantum entropy   
\beq
\label{entae}
S_{12}(\rho_A) \equiv S\left( \rho_A\, ||\, \rho_{A,1} \otimes \rho_{A,2} \right)
\eeq
where $\rho_{A,1} \otimes \rho_{A,2}$ is defined as the tensor product of the reduced density matrices
\beq
\label{entaf}
\rho_{A,1} = \Tr_{\HH_{A,2}} \left[ \rho_A \right]~, ~~
\rho_{A,2} = \Tr_{\HH_{A,1}} \left[ \rho_A \right]
~.
\eeq
This quantity vanishes if and only if our system is completely decoupled into 
the two subsystems 1 and 2. In fact, one can show that the definition \eqref{entae} is simply 
the quantum mutual information 
\beq
\label{entag}
S_{12}(\rho_A) = S(\rho_{A,1}) +S(\rho_{A,2}) - S(\rho_A)
\eeq
where $S(\rho)$ is the standard entanglement entropy \eqref{introaa} and 
$S(\rho_{A,1})$, $S(\rho_{A,2})$ are `inter-system' entanglement entropies. 
A version of the latter with $A^c = {\O}$ was studied recently in the context of holography
in \cite{Mollabashi:2014qfa}.

As a concept, separability is very well adapted to describe properties related to the connectivity
index and its behavior under changes of the system, e.g. under renormalization group flows
that lead to Hilbert space fragmentation. We will soon examine these properties in a specific 
context of large-$N$ quantum field theories.

\section{Hilbert space fragmentation in quantum field theory}
\label{fragment}

There are several common mechanisms in quantum field theory that change the connectivity index.
For example, in strongly coupled gauge theories an operator will frequently hit the unitarity bound
and decouple from the remaining degrees of freedom as a free field.\footnote{There are many 
well known examples of this type of decoupling. For instance, a class of three-dimensional 
superconformal field theories with a rich pattern of such features at strong coupling was studied 
in \cite{Morita:2011cs}.} Another common example, involves gauge theories whose gauge group 
$G$ is Higgsed. In the IR one obtains a product gauge group $G_1 \times G_2 \times \cdots G_n$. 
In both cases the Hilbert space fragments and the connectivity index (as defined in the introduction)
increases. It should be noted that there are also situations where the 
connectivity index may decrease under RG running. This occurs naturally in RG flows where a 
mass gap develops in the IR, e.g.\ a massive degree of freedom is removed from 
the spectrum in the far IR or a gauge group confines.

In this paper we will examine closely the case of gauge group Higgsing in the Coulomb branch of
large-$N$ superconformal field theories. A concrete example of the general setup has the following ingredients. 
The UV conformal field theory CFT$_{\rm UV}$ is a gauge theory with gauge group $SU(N)$. It flows by Higgsing 
to an IR conformal field theory which is a product of decoupled theories, e.g.\ CFT$_{\rm IR}=$ CFT$_1 \times\, $CFT$_2 \times\, $CFT$_3$.  
CFT$_1$ is a gauge theory with gauge group $SU(N_1)$, and CFT$_2$ is a gauge theory with gauge group $SU(N_2)$ $(N=N_1+N_2)$. 
CFT$_3$ denotes collectively a $U(1)$ gauge theory with a set of free decoupled massless fields.
The massless scalar fields in this set express the moduli whose vacuum expectation value 
Higgses the UV gauge theory and sets the vacuum state.

It is interesting to consider the low-energy effective description of this theory.
At small energies above the extreme IR the direct product theory is deformed by irrelevant interactions of three different types
\beq
\label{multiaa}
\int d^{p+1} x \Big( g_1 V_1 +g_2 V_2 +\ldots + h_{12} \OO_1 \OO_2 +\ldots 
+ \LL(\varphi,\Phi_1,\Phi_2) \Big)
~.
\eeq
The first type includes the operators $V_1$ and $V_2$, which are single-trace operators in 
CFT$_1$ and CFT$_2$ respectively (with irrelevant couplings $g_1$, $g_2$ of order $N$). 
The second type involves a double-trace operator of the form $\OO_1 \OO_2$, 
where $\OO_1$ and $\OO_2$ are single-trace in CFT$_1$ and CFT$_2$.\footnote{The 
double-trace coupling $h_{12}$ is $\OO(N^0)$. The overall Lagrangian is normalized 
so that all terms are $\OO(N^2)$.} 
The third type, $\LL(\varphi,\Phi_1,\Phi_2)$, is an interaction between the fields of CFT$_3$
(collectively denoted here as $\varphi$) and gauge-invariant composite operators (single-trace
or multi-trace) of CFT$_1$ and CFT$_2$ (denoted as $\Phi_1$ and $\Phi_2$ respectively).
For example, $\LL$ can include interactions of the form $\varphi V_1$ and $\varphi V_2$ in which
case the single-trace couplings $g_1,g_2$ become dynamical.
The dots in \eqref{multiaa} indicate interactions of higher
scaling dimension, i.e.\ more irrelevant operators, that become increasingly important as we increase
the energy.

Explicit examples of such operators and the corresponding irrelevant interactions will be provided
in the next section \ref{multicenterIR} for $\NN=4$ SYM theory. 

So far our discussion is valid at any $N$.
We notice that the non-abelian IR CFTs, CFT$_1$ and CFT$_2$, communicate directly
only by multi-trace operators, as dictated by gauge invariance 
(a point emphasized in \cite{Kiritsis:2008at}), and indirectly via the interaction with abelian 
degrees of freedom of CFT$_3$. At finite $N$ both types of interactions contribute to the precise manner  
in which the system passes from a non-separable UV state to an extreme IR separable state. 
However, in the large-$N$ limit\footnote{We consider the large-$N$ limit in  
both CFT$_1$ and CFT$_2$, $i.e.$ $N_1,N_2 \to \infty$ with the ratio $N_1/N_2$ kept fixed.} 
many of the effects of the multi-trace operators are subleading in the $1/N$-expansion. 
In particular, we provide evidence in 
section \ref{moreIR}, that the effects of multi-trace operators that break the IR separability 
are suppressed at leading order in $1/N$
and the leading effects come from the communication mediated by the $U(1)$ degrees of freedom.
As we move up in energy the irrelevant interactions become stronger and the IR effective expansion
in \eqref{multiaa} eventually resums. At some characteristic energy scale ---comparable 
to the scale set by the vacuum expectation value that Higgsed the UV gauge group--- one eventually
enters the explicitly non-separable description of the UV $SU(N)$ gauge theory.

The main purpose of this paper is to quantify this transition using the measures of entanglement
presented in the previous section \ref{entangle} and to explore potentially new features associated
with the large-$N$ limit. We will focus on large-$N$ quantum field theories with a weakly curved gravitational
dual.

\paragraph{Entanglement entropy.}
The entanglement entropy of large-$N$ conformal field 
theories with a weakly curved gravitational dual can be computed efficiently using the Ryu-Takayanagi
prescription in the AdS/CFT correspondence. This computation, which will be performed in the next four
sections, involves the analysis of a minimal co-dimension-2 surface in multi-centered brane geometries
in ten- or eleven-dimensional supergravities. The non-standard feature of this computation is that the 
minimal surface embeds non-trivially along the compact manifolds transverse to AdS. We will see that 
the above-mentioned transitions of the connectivity index are closely related to the formation 
of a separatrix in the geometry of the Ryu-Takayanagi minimal surface.

\paragraph{Relative entropy of entanglement and quantum mutual information.}
In section \ref{entangle} we presented two measures of separability, the relative entropy of entanglement
$D_{\rm REE}(\rho)$ \eqref{entad} and the quantum mutual information $S_{12}(\rho_A)$ \eqref{entag}.
Currently, we are not aware of an efficient computational method for such quantities in interacting
quantum field theories, either directly in quantum field theory or holographically. Nevertheless,
the above discussion indicates that we should anticipate the following features. 

To specify $S_{12}(\rho_A)$ we define subsystem 1 as the subsystem associated with the degrees 
of freedom of the IR CFT$_1$. The subsystem 2 (that refers to the complementary Hilbert space) 
includes the remaining degrees of freedom of the full $SU(N)$ theory. In the effective IR description 
subsystem 2 includes the degrees of freedom of CFT$_2$ and CFT$_3$.
Since we are considering a non-trivial RG flow the relative quantum entropy on a sphere of radius $\ell$
will be a non-trivial function of $\ell$. Complete decoupling in the extreme IR implies
that $S_{12}(\ell)$ vanishes at $\ell=\infty$ and increases as $\ell$ decreases 
towards $\ell=0$ (that probes the extreme UV). The increasing positive magnitude of 
\beq
\label{multiaba}
S_{12}(\rho_A) = S(\rho_A) - S(\rho_{A,1}) - S(\rho_{A,2})
\eeq
is a measure of the increasing strength of correlation of the degrees of freedom of the IR CFT$_1$
with the rest of the system at high energies. The general discussion in the beginning of this section
suggests that this increase is suppressed in the large-$N$ limit at low energies because the effects of 
inter-system interactions mediated by multi-trace operators are suppressed. It is of interest to 
understand if this expectation is verified by the explicit computation of $S_{12}(\ell)$, and 
to determine precisely how $S_{12}(\ell)$ interpolates between 
the extreme UV and IR descriptions that exhibit a different connectivity index.
We anticipate a qualitatively similar behavior from other measures of separability,
e.g.\ the relative entropy of entanglement $D_{\rm REE}(\rho)$. 
An efficient computational method for the relative quantum entropy would be helpful
in addressing these issues, but goes beyond the immediate goals of this paper.

\section{Coulomb branch of SCFTs and multi-centered geometries}
\label{multicenter}

In this preparatory section we collect useful facts and notation for the geometries involved in 
the holographic computation of the entanglement entropy in the Coulomb branch of superconformal
field theories.

\subsection{Notation and main features of multi-centered brane geometries}

We focus on supersymmetric conformal field theories with a weakly curved gravitational dual
in string/M-theory. The gravitational description of the Coulomb branch of these theories is directly 
related to the geometry of a discrete collection of flat parallel D/M-branes in 10 or 11-dimensional supergravity. 
This geometry is uniquely specified by a single harmonic function $H=H(\vec{y})$, where $\vec{y}$ 
are the coordinates transverse to the brane volume. The supergravity solution also carries charge under the 
corresponding $(p+1)$-form gauge fields and generically sources the dilaton 
$\Phi$.\footnote{The specific well known expressions for these fields can be found in the literature. 
Here we will concentrate on the metric, which is the only field that participates directly in our computation.}

In this paper, we will focus on multi-centered geometries given by: 
\begin{itemize}
\item[$\bullet$] D3 branes in $D=10$ dimensions, relevant for the $d=4$ $\NN=4$ SYM theory,
\item[$\bullet$] M2 branes in $D=11$ dimensions, relevant for the $d=3$ $\NN=8$ ABJM Chern-Simons-Matter
theories \cite{Aharony:2008ug},
\item[$\bullet$] D1-D5 bound states in $\mathbb{R}^{1,5}\times \mathcal{M}^4$. The D5 branes
wrap the compact manifold $\MM_4$ (usually taken as $\mathbb{T}^4$ or $K3$) and give rise at low energies to an
interacting $(1+1)$-dimensional superconformal field theory.
\end{itemize}

\noindent
The corresponding geometries in asymptotically flat space\footnote{The metric of the D3 and D1-D5
systems is given here in the string frame of type IIB string theory.} are given by the metrics
\bea
\mathrm{D3}:& &
ds^2= H_3^{-1/2} \eta_{\mu\nu}dx^{\mu}dx^{\nu}\, +\, H_3^{1/2} \delta_{ij}dy^i dy^j\ ,                               
\rule{.0pt}{.5cm}        \label{D3MULTI}\\
\mathrm{M2}:& &
ds^2= H_2^{-2/3} \eta_{\mu\nu}dx^{\mu}dx^{\nu}\, +\, H_2^{1/3} \delta_{ij}dy^i dy^j\ ,                              
\rule{.0pt}{.7cm} \label{M2MULTI}\\
\rm{D1D5}:& & ds^2=\left( H_1 H_5 \right)^{-1/2}\eta_{\mu\nu}dx^{\mu} dx^{\nu} + 
\left( H_1 H_5 \right)^{1/2} \delta_{ij} dy^i dy^j + \left( \frac{H_1}{H_5}\right)^{1/2} ds^2(\mathcal{M}^4).\ 
\rule{.0pt}{.7cm}  \label{D1MULTI}
\eea

The harmonic functions $H_3$, $H_2$ are
\bea
H_3(\vec{y})=1+\sum_{I=1}^K \frac{N_I \rho_3}{ |\vec{y}-\vec{y}_I |^4 }~,&\rule{.5cm}{.0pt} &
\rho_3= 4\pi g_s \alpha'^2 \label{MULTIC} \\
H_2(\vec{y})=1+\sum_{I=1}^K \frac{M_I \rho_2}{ |\vec{y}-\vec{y}_I |^6 }~,& &\rho_2= 2^5 \pi^2 \ell_P^6
\eea 
The vectors $\vec{y}_I$ locate the position of the different stacks of branes in the transverse space. We are 
considering $K$ stacks of D3 (M2) branes, each one made out of $N_I$ D3 branes ($M_I$ M2 branes). 
$g_s$ and $\alpha'$ are the string coupling constant and string Regge slope. $\ell_P$ is the 
eleven-dimensional Planck length.

For the D1-D5 system, the two harmonic functions $H_1$ and $H_5$ are:
\bea
H_1(\vec{y})= 1+\sum_{I=1}^K \frac{Q^{(1)}_I \rho_1}{|\vec{y}-\vec{y}_I |^2 }~, &\rule{.5cm}{.0pt}  & 
\rho_1= \frac{g_s \alpha'}{v}\\ 
H_5(\vec{y})= 1+\sum_{I=1}^K \frac{Q^{(5)}_I \rho_5}{|\vec{y}-\vec{y}_I |^2 }~, & & \rho_5=g_s \alpha'
\eea
where $v$ is essentially the volume of $\mathcal{M}_4$, i.e.\ $v={V_4}/(2\pi)^4\alpha'^2$.
It will be technically convenient to focus on D1-D5 bound states with parameters that obey the relation
\beq\label{RATIOD1D5}
\frac{Q^{(1)}_J}{Q^{(1)}_1} = \frac{Q^{(5)}_J}{Q^{(5)}_1}  \qquad \forall\  1<J\le K\ .  
\eeq
This restriction guarantees that the dilaton $\Phi$, given by the relation $e^{2\Phi}= H_1/H_5$,
will be constant in the near-horizon limit.

\smallskip
\paragraph{Near-horizon limit.}

For the D3 and D1-D5 branes, the decoupling limit \cite{Maldacena:1997re} is defined by 
sending $\alpha'\to0$, and keeping the ratios\, $\vec{u}=\vec{y}/\alpha'$\, and\, 
$\vec{u}_I=\vec{y}_I/\alpha'$\, fixed. As a result, the $1$ in the harmonic functions drops 
out, and the geometry remains finite in units of $\alpha'$. Under the assumption 
(\ref{RATIOD1D5}), the product $H_1 H_5$ simplifies
\beq\label{MALDAD1D5}
H_{1\cup5}\equiv(H_1 H_5)^{1/2}= \sum_{I=1}^K  \frac{Q_I \rho_{{1\cup5}} }{|\vec{u}-\vec{u}_I |^2 }\ ,\qquad 
Q_I= \sqrt{Q^{(1)}_IQ^{(5)}_I}\ ,\qquad \rho_{1\cup5}= \frac{g_s^2\alpha'^2}{v}\ .
\eeq
For the M2 branes the decoupling limit is obtained by sending $\ell_P\to 0$ and keeping\,
$\vec{u}=\vec{y}/\ell_P^{3/2}$\, and\, $\vec{u}_I=\vec{y}_I/\ell_P^{3/2}$  fixed. 

In summary, the D1-D5 system is now described by the function $H_{1\cup5}$, and the D3 and M2 
backgrounds are described by
\beq
\label{MALDAD3M2}
H_3\,\to\, \sum_{I=1}^K \frac{N_I \rho_3}{|\vec{u}-\vec{u}_I |^4 }\, ,\qquad 
H_2\, \to\, \sum_{I=1}^K \frac{M_I \rho_2}{|\vec{u}-\vec{u}_I |^6 }
~.
\eeq

The resulting geometry interpolates between an $AdS_{p+2}\times S^q$ space at 
$|\vec{u}|\to\infty$, that captures the UV fixed point with connectivity index 1, and a decoupled product of 
$K$ $AdS_{p+2}\times S^q$ spaces as $\vec u$ gets scaled towards the centers $\vec{u}_I$. The latter 
describes the extreme IR fixed point with connectivity index $K$.

\subsection{UV physics}
\label{multicenterUV}

For the cases we analyze the asymptotic $|\vec{u}|\to\infty$ geometry is an $AdS_{p+2}\times S^q$ 
space with $(p,q)=(1,3), (2,7), (3,5)$ for the D1-D5, M2 and D3 brane systems respectively. 
The radius of each $AdS_{p+2}$ space is 
\bea
{\rm D}3:&\qquad & R_{UV}^2=\left( 4\pi g_s\, \sum_{I} N_I \right)^{1/2}\label{UVRADIAD3}\\
M2:&\qquad & R_{UV}^2= \frac{1}{4}\left(2^5\pi^2\ell_P^6\, \sum_I M_I \right)^{1/3} \label{SPHERERADIUS}\\
D1D5:&\qquad & R_{UV}^2= \left( \frac{g_s^2}{v} \sum_I Q_I \right) 
~.
\label{UVRADIA}
\eea

These UV $AdS_{p+2}\times S^q$ geometries are dual to the 
microscopic $(p+1)$-dimensional superconformal field theories mentioned in the beginning of the previous 
subsection. For concreteness, let us focus for the moment on the most emblematic case, i.e.\ the duality between 
string theory on $AdS_5\times S^5$ and $\mathcal{N}=4$ $SU(N)$ SYM theory.

The multi-centered D3 brane solutions are dual to a configuration in $\mathcal{N}=4$ SYM in which 
the $SU(N)$ gauge group has been Higgsed down to 
$SU(N_1)\times\ldots\times SU(N_K) \times U(1)^{K-1}$
($N=N_1+\ldots+N_K$). Conformal invariance, as well as the $SO(6)$ R-symmetry of the theory,
are broken by the non-vanishing vacuum expectation value of the gauge-invariant chiral operators 
\beq\label{CHIRAL}
\mathcal{O}^{(n)}\,\propto\, C_{i_1,\ldots,i_n}^{(n)}  \mathrm{Tr}\big[ X^{i_1}\ldots X^{i_n}\big] 
~,
\eeq
where $C_{i_1,\ldots,i_n}^{(n)}$ are totally symmetric traceless rank $n$ tensors of the 
$SO(6)$-charged real adjoint scalars $X^i$ of the theory. These modes arise in the gravity dual from 
a Kaluza-Klein decomposition of the transverse $S^5$ space. By analyzing the asymptotic, large $|\vec u|$, 
behavior of these modes in the multi-centered geometry one can determine the vacuum expectation 
value of the operators \eqref{CHIRAL}\cite{Kraus:1998hv,Klebanov:1999tb,Skenderis:2006uy}.

\subsection{IR physics}
\label{multicenterIR}

The decoupled product of gauge theories that arises in the extreme infrared of the Coulomb branch 
translates, in the dual multi-centered geometry, into a decoupled product of $K$ string theories on 
the $AdS^{(I)}_{p+2}\times S^q_{(I)}$ spacetimes. Each of these spacetimes arises from the full multi-centered
geometry by taking the limit $\vec{u}\to\vec{u}_I$ that isolates the gravitational dynamics near the $I$-th 
center. The radius of the $I$-th $AdS$ spacetime is weighted by the single coefficient $N_I$, $M_I$ or $Q_I$, 
respectively. The $K-1$ $U(1)$ factors are decoupled sectors of singleton degrees of freedom 
that reside on the common holographic boundary of the $AdS$ spacetimes.

As explained in section \ref{fragment}, in the IR description of the RG flow the non-abelian
IR CFTs interact off criticality via an infinite set of irrelevant multi-trace interactions, and via 
irrelevant interactions mediated by the abelian singleton degrees of freedom ---the Lagrangian 
$\LL$ in equation \eqref{multiaa}.
For example, in the case of $\NN=4$ SYM theory the leading single-trace operator $V_I$
for the $I$-th non-abelian IR theory (see equation \eqref{multiaa}) is a dimension 8 operator 
of the form \cite{Intriligator:1999ai,Costa:1999sk,Costa:2000gk}
\beq
\label{IRaa}
V = \Tr\left[ F_{\mu\nu} F^{\nu\rho} F_{\rho\sigma} F^{\sigma \mu} 
-\frac{1}{4} \left( F_\mu F^{\mu\nu} \right)^2 \right] 
+\ldots
~.
\eeq
The coefficient $g_I$ 
is proportional to the sum
\beq
\label{IRab}
g_{I} \propto \sum_{J\neq I} \frac{N_J \rho_3}{|\vec u_J -\vec u_I|^4}
~.
\eeq
Note that \eqref{IRaa} is also the type of interaction that appears in the 
small field strength expansion of the 
Dirac-Born-Infeld action that describes the exit from the near-horizon throat. In the current 
context the single-trace interaction \eqref{IRaa} describes how the throat in question connects with 
the rest of the geometry.

Besides the single-trace operator \eqref{IRaa} there are also double-trace dimension 8 operators 
of the form \cite{Intriligator:1999ai}
\beq
\label{IRac}
\Tr_I \left[ F_{\mu\nu}F^{\mu\nu} \right] 
\Tr_J \left[ F_{\mu\nu}F^{\mu\nu} \right] 
+\ldots
\eeq
which mediate the direct inter-CFT interactions mentioned in equation \eqref{multiaa}.

Finally, there are interactions of the non-abelian degrees of freedom with the abelian singleton 
degrees of freedom. Part of the singleton degrees of freedom are the massless scalar fields $\vec \varphi_{I}$
associated with the $6(K-1)$ moduli $\vec u_I -  \vec u_{I+1}$. Expanding \eqref{IRab} around the 
values of the given vacuum state produces irrelevant single-trace interactions of the form
\beq
\label{IRad}
\sum_{J\neq I} \frac{\sum_{K=I}^{J-1}\vec \varphi_K \cdot (\vec u_I- \vec u_J)}
{|\vec u_I -\vec u_J|^6} V_I
~.
\eeq
This makes the single-trace couplings $g_I$ dynamical. 

Holographically, in this description we are working in the bulk with an explicit UV cutoff and we are 
dealing with a set of UV-deformed $AdS$ gravity theories coupled in two ways: by mixed boundary 
conditions and by explicit boundary degrees of freedom (the singletons) that 
make the sources of some of the bulk fields dynamical. A similar picture of coupled throat geometries
was proposed some time ago in \cite{Dimopoulos:2001ui,Dimopoulos:2001qd}.

\section{Holographic Entanglement entropy}
\label{HEEgeneric}

In a field theory in $p+1$ dimensions, the static entanglement entropy of a space-like region $A$ 
is defined as the von-Neumann entropy of the density matrix $\rho_A$ which is obtained by tracing out 
the degrees of freedom in the complement of $A$ (see equations \eqref{introaa}, \eqref{introab}). 

For conformal theories living on the boundary of $AdS_{p+2}$, the \RT\ prescription (RT) computes the 
holographic entanglement entropy (HEE) by considering the area of a $p$-dimensional minimal surface in 
$AdS_{p+2}$, whose boundary is $\partial A$. We will refer to this surface as $\surfRT$  \cite{Ryu:2006bv}. 
There is a beautiful derivation of the correctness of this prescription for spherical entangling surfaces. 
By conformally mapping the density matrix $\rho_A$ to a thermal density matrix, the authors of \cite{Casini:2011kv} 
showed that the thermal entropy of the dual hyperbolic black hole coincides with the HEE computed \`a la 
Ryu-Takayanagi. The relation between the entropy of $\rho_A$ and the minimal area condition was further 
investigated and clarified in \cite{Lewkowycz:2013nqa}.  

For non-conformal theories with a gravity dual, a natural extension of the \RT prescription was given in 
\cite{Nishioka:2006gr,Klebanov:2007ws}. These authors considered the functional
\beq\label{PRESCRIP}
S[\partial A]=\frac{1}{4G_N^D} \int d^{D-2}\xi\, e^{-\Phi}\sqrt{ \mathrm{det}\, g_{ind} }
\eeq
where $g_{ind}$ is the induced metric of a minimal co-dimension-2 surface $\surf$ in the full string theory or 
M-theory background. The surface $\surf$ is again specified to have $\partial A$ as its boundary. 
This generalized prescription is the prescription we will apply in the computation that follows. 
In our setup, the dilaton field $\Phi$ is a constant for all the cases we will consider; the D3, M2 and D1-D5 
branes. This statement is obvious for D3 and M2 branes, and follows from the assumption (\ref{RATIOD1D5}) 
in the case of the D1-D5 bound states. 

It is clear that for $AdS_{p+2}\times S^q$ spaces, the \RT prescription is in perfect agreement with 
(\ref{PRESCRIP}). When there is no dependence on the transverse sphere, the problem of a minimal
surface $\surf$ that wraps $S^q$ reduces to the problem of finding $\surfRT$ in $AdS_{p+2}$. 
The Newton constant in $AdS_{p+2}$ is related to $G_N^{D}$ through the formula 
\beq
G_N^{p+2}=G_N^{D}/\mathrm{Vol}(S^q).
\eeq
A typical class of examples in which the prescription (\ref{PRESCRIP}) is non-trivial are the confining backgrounds 
of \cite{Witten:1998zw, Klebanov:2000hb}, for which the entanglement entropy was studied in 
\cite{Klebanov:2007ws}. These backgrounds are of the type $\mathcal{M}_{p+1}\times \mathcal{C}_{D-p-1}$, 
where $\mathcal{C}$ is a cone over a certain compact manifold $\mathcal{S}$. The volume of $\mathcal{S}$ may 
shrink along the radial coordinate of the cone, and since $\gamma$ wraps $\mathcal{S}$, it will be sensitive 
to the dynamics of these extra dimensions along the RG flow.

Similarly, the multi-centered geometries of interest in this paper are not product spaces globally. They become 
locally $AdS_{p+2}\times S^q$ spacetimes only in certain asymptotic regions. If the dimension of $\surf$ 
was different from $D-2$, other data would be needed to determine it, and the surface would not be unique 
for a given $\partial A$. An example appears in the holographic computation of the Wilson loop in 
\cite{Minahan:1998xq}.

\subsection*{Multi-centered geometries} 
\label{HEEMULTICENTER}

The remainder of this section provides a qualitative description of the surface $\gamma$ in the 
multi-centered backgrounds described previously. We consider spherical entangling 
surfaces when $p=2,3$, and intervals when $p=1$. It is useful to choose space-like coordinates 
adapted to these geometries. In dimensions $p=2,3$, we choose spherical coordinates: 
$\vec{x}=(\rad,\phi_1,\ldots,\phi_{p-1})$, where $\rad>0$ is the radius of the sphere and 
$\vec{\phi}$ are angles. In one dimension we use a similar notation: $\rad$ is the spatial field
theory coordinate that runs along the real line. The entangling region $A$ is described by the equation 
$\rad^2<\eLL^2$. This means that $\rad\in\mathcal{I}_{\eLL}$ where $\mathcal{I}_\eLL=(-\eLL,\eLL)$ for $p=1$, 
and $\mathcal{I}_\eLL=[0,\eLL)$ for $p=2,3$. 

The main example we will consider in detail is the case of the two-centered geometry. The two-centered 
geometries are conveniently described by hyper-cylindrical coordinates in the transverse space. 
The branes are separated along a direction $z$, and the space orthogonal to $z$ is described by hyper-spherical 
coordinates $(y,\Omega_{1},\ldots,\Omega_{q-1})$. In this setting, the functions $H_i$ of the previous 
section will depend both on $z\in\mathbb{R}$ and $y>0$. The origin $z=0$ is taken to be the center of mass. 
We can also introduce polar coordinate in the $(z,y)$ plane,  
$$
z=r \cos\theta, ~~y=r\sin\theta
$$ 
with $r>0$ and $\theta\in[0,\pi]$. For coincident branes, $K=1$, the coordinate $r$ becomes the radial 
coordinate of $AdS_{p+2}$, and $\theta$ becomes the polar angle of the $q$-sphere.

The minimal surface is static with Dirichlet boundary conditions in the time direction, which will not play 
any further role.  The coordinates describing the co-dimension-2 surface are chosen as follows
\bea
\xi_i&=&\phi_i~,\ \ \ i=1,\ldots,p-1,  \nn \\ 
\xi_{j+p-1}&=&\Omega_j~,\ \ j=1,\ldots, q-1, \nn \\
\xi_{D-3}&=&\theta , \nn\\
\xi_{D-2}& =&\rad . \label{COORD}
\eea
The embedding in the $D$-dimensional background is specified by the function $r(\rad,\theta)$, where 
$\rad\in A$ and $\theta\in[0,\pi]$. This function is an interesting object because it mixes the evolution along a 
field theory direction, $\rad$, with the change of the geometry along the transverse space direction $\theta$. 
The non-trivial dependence on $\theta$ originates from $H_i$ which are explicit functions of $\theta$. 

The behavior of $r(\rad,\theta)$ can be understood qualitatively by regarding $r(\rad,\theta)$ as a map from 
$\mathcal{I}_\eLL\times [0,\pi]$ to the plane $(z,y)$. We imagine foliating the surface $r(\rad,\theta)$ by fixing a 
certain $\radz$, drawing the curve $\mapz(\theta)=r(\radz,\theta)$ in the plane $(z,y)$, and moving $\radz$ in the 
interval $\mathcal{I}_\eLL$. For example, in $AdS_{p+2}\times S^q$, the solution is given by the \RT surface 
which is $\theta$ independent, therefore $r(\rad,\theta)=r(\rad)$, and the map $\mapz(\theta)$ draws circles of 
radius $r(\radz)$. From this simple analysis we are able to infer three out of the four boundary conditions that fix a 
generic $\theta$-dependent solution on $\mathcal{I}_\eLL\times [0,\pi]$:
\beq\label{BD3}
r(\rad,\theta)\Big |_{\sigma=\eLL}=\infty\, ,\qquad \partial_{\theta} r(\rad,\theta)\Big|_{\theta=0}=0\, 
,\qquad \partial_{\theta} r(\rad,\theta)\Big|_{\theta=\pi}=0\, .
\eeq
We will discuss the boundary condition at $\rad=0$ in a moment.

To start thinking about $r(\rad,\theta)$ in two-centered solutions, it is useful to first consider the limit $\eLL\to\infty$.
In this limit the surface probes the physics of the deep IR of the field theory where the UV gauge group has been 
Higgsed and the energy scales of interest are well below the mass of the massive $W$ bosons. 
In the gravity dual this limit zooms into the vicinity of the two centers which can be regarded as decoupled.
The surface $\surf$ is then given by the union $\surf_1\cup\surf_2$, where $\surf_i = \surfRT\times S^q$. 
At this point, it is important to recall that $\surfRT$ has a turning point at $\sigma=0$, i.e.\ $r(\rad)>r(0)$ for any 
$\rad\in\mathcal{I}_\eLL$. The fact that $\rad=0$ is the turning point follows from the 
symmetries of the entangling surface and from the assumption that $\surfRT$ is convex. 

\begin{figure}[t]
	\centering
		\begin{tikzpicture}

			\filldraw[fill=green!25,draw=green!50!black] (1,1) rectangle (5,5);  
			\node[scale=1.1] at (.7,1.2) {$0$};
			\node[scale=1.1] at (1.2,.7) {$0$};
			\node[scale=1.4] at (.7,4.8) {$\pi$};
			\node[scale=1.2] at (5,.7) {$\eLL$};
			\draw[-triangle 45, black, line width=.1pt] (1.4,.68)--(4.6,.68);
			\node at (3,.4) {$\rad $};
			\draw[blue,dashed,line width=1pt] (4.5,1)--(4.5,5);
			\draw[blue,dashed,line width=1pt] (4.75,1)--(4.75,5);
		
			\node[scale=1] at (4.5,5.3) {$-\epsilon$};
			
			\draw[red,very thick] (2.5,1)--(2.5,5);
			\draw[brown,very thick] (1.25,1)--(1.25,5);
			\draw[brown,very thick] (1.5,1)--(1.5,5);
			\draw[brown,very thick] (1.75,1)--(1.75,5);
		       \draw[green!50!black, line width=1pt] (1,1) rectangle (5,5); 
		       
			\draw[-triangle 45, black,line width=.6pt] (6.2,1.5)--(15.3,1.5);
			
			\draw[-triangle 45, black,line width=.1pt] (13.9,5)--(14.9,5);
			\draw[-triangle 45,black,line width=.1pt] (14,4.9)--(14,6);
			\node[scale=.9] at (14.9,4.8) {$z$};
			\node[scale=.9] at (13.8,5.9) {$y$};
			
			\filldraw [black]  (12,1.5) circle (2pt)
			                          (9,1.5) circle (2pt);
		
                      \node[scale=1] at (9,1.1) {$z_1$};
		        \node[scale=1] at (12,1.1) {$z_2$};
		
			\draw[brown,very thick] (8.75,1.5) .. controls (8.75,2) and (9.25,2) .. (9.25,1.5);
			\draw[brown,very thick] (8.5,1.5) .. controls (8.5,2.2) and (9.5,2.2) .. (9.5,1.5);
			\draw[brown,very thick] (8.25,1.5) .. controls (8.25,2.4) and (9.75,2.4) .. (9.75,1.5);
			
			\draw[brown,very thick] (11.75,1.5) .. controls (11.75,2) and (12.25,2) .. (12.25,1.5);
			\draw[brown,very thick] (11.5,1.5) .. controls (11.5,2.2) and (12.5,2.2) .. (12.5,1.5);
			\draw[brown,very thick] (11.25,1.5) .. controls (11.25,2.4) and (12.75,2.4) .. (12.75,1.5);
			
			\draw[blue, dashed, very thick] (6.8,1.5) .. controls (6.8,6.3) and (14.7,6.3) .. (14.7,1.5);
			\draw[blue, dashed, very thick] (6.6,1.5) .. controls (6.6,6.5) and (14.9,6.5) .. (14.9,1.5);
			

			\draw[-triangle 45,black,line width=.6pt] (10.8,1.)--(10.8,6.5);
			\draw[red, very thick] (7.6,1.5) .. controls (7.6,3.5) and (10.8,4) .. (10.8,1.5);
			\draw[red, very thick] (10.8,1.5) .. controls (10.8,4) and (13.4,3.5) .. (13.4,1.5);
			
			\draw[-triangle 45, black,line width=.1pt] (2.8,2.5) .. controls (4,5) and (6,4) .. (8,3);
			\node[scale=.95] at (6,4.3) {$r(\rad,\theta)$};
			\node[scale=.95] at (8.8,5.3) {$AdS_{UV}\times S^q$};
		
		\end{tikzpicture}
		\caption{\it \small Qualitative behavior of the map $\mapz(\theta)$ as a function of $\radz$, and for large 
		values of $\eLL$. The red line represents the separatrix. Below the separatrix, a suitable set of variables  
		that describe the surface will be given in Section \ref{SEC_COORD}. }
		\label{FIGURA}
\end{figure}
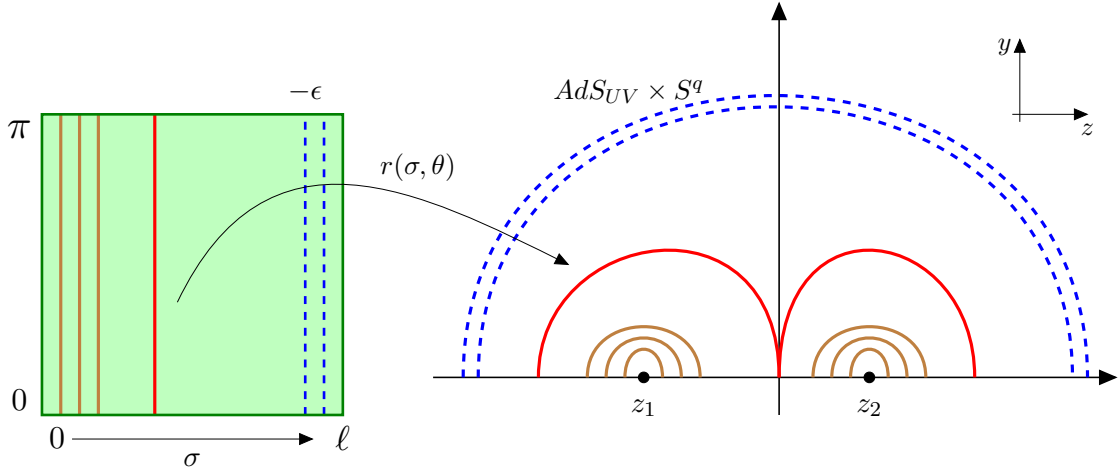

When $\eLL$ is finite, but large enough for $\surf$ to probe the IR throats, the picture we have just described 
will be approximately valid only locally close to each of the two centers. In a neighborhood of $\radz=0$ the 
map $\mapz(\theta)$ draws approximately small disconnected circles around the position of each stack of 
branes (points $z_1$ and $z_2$ in the $(z,y)$ plane in Figure \ref{FIGURA}). The curve 
$r_m(\theta)\equiv r(\rad=0,\theta)$ generalizes the notion of turning point in the $AdS$ \RT surface
and obeys the boundary conditions
\beq\label{BD4}
 \partial_\rad r(\rad,\theta)=0\quad \mathrm{at} \quad \rad=0 \quad \mathrm{for\ any} \ \theta. 
\eeq
The overall picture in the IR is summarized by the brown curves in Figure \ref{FIGURA}.

The above description refers to the IR patch of the surface $\gamma$ associated to a space-like region of 
large enough radius $\eLL$. In the opposite regime, we can ask what happens at 
$\radz=\eLL-\epsilon$ ($\epsilon \ll \eLL$), when the curve $r(\eLL-\epsilon,\theta)$  is close to the UV 
boundary. Because the boundary is $AdS_{UV}\times S^q$, this curve is again approximately 
$\theta$-independent and the associated map $\mapz(\theta)$ draws a large circle in the $(z,y)$ plane 
(captured by the blue curves in Figure \ref{FIGURA}).  

The inevitable conclusion of the above analysis is that, although the surface $\gamma$ is always
simply connected, the topology of the curves 
$\big\{\mapz(\theta)\big\}_{\radz\in\mathcal{I}_\eLL}$ may change as we vary $\sigma_0$. 
When $\eLL$ is large enough, the minimal surface will have a UV patch where $\mapz(\theta)$ is 
topologically $S^1$, and an IR patch where $\mapz(\theta)$ is topologically $S^1\times S^1$. 
For such a surface $\surf$ there is necessarily a branch point. The curve $r_b(\theta)\equiv r(\rad_b,\theta)$ 
at which this branch point belongs will be referred to from now on as the {\it separatrix}. This is sketched as 
the red line in Figure \ref{FIGURA}. 

The topology change that we described above does not occur for surfaces with small enough $\ell$
that can only probe the UV part of the full geometry. For such surfaces the curves $\mapz(\theta)$ are 
topologically $S^1$ for any $\radz\in\mathcal{I}_{\eLL}$. It is clear that the discriminating quantity between 
the existence of the topology change or not, for a given $\ell$, is the turning point curve $r_m(\theta)$. 
Accordingly, we will distinguish between the following two phases:
\begin{itemize}
\item[$\bullet$] Phase \UNO,\ for $\eLL<\eLL_{c}$, where the topology of $r_m(\theta)$ is $S^1$. 
In this case we can describe $\surf$ with single-valued coordinates.  
\item[$\bullet$] Phase \DUE,\ for $\eLL>\eLL_{c}$, where the topology of $r_m(\theta)$ is $S^1\times S^1$. 
In that case a \separatrix exists and when $\mapz(\theta)$ moves below the separatrix, $r(\rad,\theta)$ 
becomes double-valued. 
\end{itemize}

The counterpart of the transition between these phases in field theory is a transition of the behavior of 
the entanglement entropy as a function of $\ell$ at $\ell_c$.

The qualitative behavior of $\surf$ for multi-centered geometries can be deduced by following the 
same logic as in the two-centered solution. However, in the general case it will not be possible to 
restrict the discussion to a certain plane $(z,y)$, and one has to consider the full transverse space.

\section{UV expansion of the entanglement entropy}
\label{HEE2uv}

In this section we will study more explicitly the HEE of phase \UNO. The equation of motion of 
$r(\rad,\theta)$ is a non-linear, quite challenging, PDE. Yet, we are able to obtain a series 
expansion of the solution by expanding in a small dimensionless parameter that combines the 
mass scale of symmetry breaking (equivalently the center separation in the geometry) and the 
sphere radius $\ell$. Our perturbative solution is analytic in the variables $\rad$ and $\theta$, 
and at zeroth order coincides with the AdS $\surfRT$ solution. The perturbative solution does not allow
us to detect analytically the formation of the separatrix as we approach $\ell_c$, but it confirms the 
qualitative description of the previous section.

By direct integration of the generalized HEE functional we obtain a series of finite corrections to the 
$AdS_{p+2}$ entanglement entropy. Perhaps suprisingly, the translation of the result to field
theory language suggests that the lowest chiral primary operators do not contribute to these corrections. 

\subsection{Minimal surface action and its equations of motion}

In phase \UNO\ the variable $r(\rad,\theta)$ is single-valued as a function of $\theta$, thus we can 
write the induced metric on $\surf$ by using the coordinates (\ref{COORD}). Referring to the 
components of the background metrics (\ref{D3MULTI}), (\ref{M2MULTI}) and (\ref{D1MULTI}), 
with the generic notation, 
\bea
ds^2=g_{\mu\nu}dx^{\mu} dx^{\nu} \nn ~, ~~
\vec{x}=(t,\rad,\phi_1,\ldots,\phi_{p-1},r,\theta,\Omega_{1},\ldots,\Omega_{q-1})\ , 
\eea
the induced metric on $\surf$, in the coordinates (\ref{COORD}), is given by
\be
ds^2_{ind}= ds^2_{ind}\Big|_{(\rad,\theta)} + ds^2_{ind}\Big|_{(\phi,\Omega)}\ ,
\ee 
where
\bea
ds^2_{ind}\Big|_{(\rad,\theta)} &=&
\left( g_{\rad\rad}+ g_{rr} \left(\frac{\partial r}{\partial\rad}\right)^2 \right)d\rad^2 
+2 g_{rr} \frac{\partial r}{\partial\rad}\frac{\partial r}{\partial\theta}d\rad d\theta + 
\left( g_{\theta\theta} + g_{rr}\left(\frac{\partial r}{\partial\theta} \right)^2\right)d\theta^2 \nn ~,\\ 
ds^2_{ind}\Big|_{(\phi,\Omega)} &=&\, g_{ij}d\phi^i d\phi^j\, +\, g_{ab}d\Omega^a d\Omega^b\ .
\eea
The HEE functional is then 
\beq\label{ACTIONUV}
S_p=\frac{1}{4G_N^D}\, \int d\vec{\Omega}\, \sqrt{g_{ab}}\, 
\int d\vec{\phi}\, \sqrt{g_{ij}}\, \int d\rad d\theta\, \mathcal{L}_p\left[ \theta,r(\rad,\theta) \right]
\eeq
where the Lagrangian $\mathcal{L}_p$ can be put into a form valid for all cases of interest here 
(the D3, M2 and D1-D5 branes), 
\beq\label{UVLAGR}
\mathcal{L}_p= \, \rad^{p-1}\, \mathcal{K}[\theta,r]\, \mathcal{H}[\theta,r]\, 
\sqrt{1+ \frac{\partial_{\theta}r^2}{r^2} + 
\left(\mathcal{H}[\theta,r]\right)^2\, \partial_\rad r^2}
~.
\eeq
In (\ref{UVLAGR}) we defined the functions
\beq
\begin{array}{cllll}
D3:\qquad & &\displaystyle{ \mathcal{H}^2=H_3}~, & & \displaystyle{ \mathcal{K}={r^5\sin^4\theta}} \\ 
M2:\qquad & &\displaystyle{\mathcal{H}^2= H_2}~,& &  \displaystyle{ \mathcal{K}={r^7\sin^6\theta} }
\rule{.0pt}{.8cm}\\
D1D5:\qquad &\rule{.6cm}{.0pt} &\displaystyle{\mathcal{H}^2= {H^2_{1\cup 5}}}~,& \rule{.5cm}{.0pt} 
&  \displaystyle{\mathcal{K}=r^3\sin^2\theta }\rule{.0pt}{.8cm}
~.
\end{array}
\eeq

In the two-centered geometries we fix the origin of the $z$ axis at the center of mass of the system,
namely we set
\beq\label{CM}
z_1 N_1+ z_2 N_2=0
~. 
\eeq
After the implementation of the condition (\ref{CM}), the Euler-Lagrange equation following from 
(\ref{UVLAGR}) depends only on a single dimensionful parameter, $z_1$ for example. 
Schematically, the single PDE that we need to solve is the equation of motion of $r$
\beq
\label{PDEuv}
\mathrm{Eq}\big[r(\rad,\theta), z_1\big]=0
~.
\eeq
The explicit form of this equation is provided in Appendix \ref{APPEQ}.

\subsection{Perturbative UV Solution} 
\label{PERTURBSECTION}

Before entering the details of the calculation, we review the $AdS$ solution making manifest the 
underlying scale invariance. This is our starting point towards a perturbative solution of the 
non-linear PDE \eqref{PDEuv} that follows from (\ref{UVLAGR}).

It is convenient to work with the variable $\zeta=1/r^2$, in the cases of D3 and D1-D5 branes, and 
$\zeta=16 R^2/r^4$  for the M2 branes. The UV boundary is now at $\zeta=0$.  In our conventions, the 
metric of $AdS_{p+2}$ is written as
\beq
ds^2= \frac{1}{R^2}\, \frac{1}{\zeta}\left( -dt^2+ d\rad^2 + \rad^2 d\vec{\phi}_{p-1}^{\,2}
+ R^4 \frac{d\zeta^2}{4\zeta} \right)\, , \label{UVADSUV}
\eeq
where $R=R_{UV}$ is the $AdS$ UV radius defined case by case in \eqref{UVRADIAD3}-\eqref{UVRADIA}. 
The \RT surface is obtained from the embedding function $\zeta(\rad)$. Its equation of motion and 
the corresponding solution are,
\bea
\mathrm{Eq}\big[\zeta(\rad),z_1=0\big]&=& \zeta''+\frac{p-1}{2}\frac{\zeta'^2}{\zeta}
+\frac{p-1}{x}\zeta' \left(1+\frac{R^4}{4}\frac{\zeta'^2}{\zeta} \right)+\frac{2p}{R^4}\,=\, 0\, , \label{EQF0}\\
\zeta(\rad)&=& \frac{\eLL^2}{R^4}\left(1- \frac{\rad^2}{\eLL^2} \right)
\equiv \frac{\eLL^2}{R^4}\,\Func\left( \frac{\rad}{\eLL} \right)\, . \label{ADS_SOL}
\eea
It should be noted that with our choice of spherical entangling surfaces, the embedding function is 
independent of $p$. In the r.h.s.\ of (\ref{ADS_SOL}) we wrote $\zeta(\rad)$ in a conformal fashion:  
we isolated the pre-factor $\ell^2$, and defined the function $F(\yc)$ that depends only on the dimensionless 
combination $\yc=\rad/\eLL$. The pre-factor captures the weight of $\zeta(\rad)$ under rescaling of $\ell$. 
We also notice that the equation (\ref{EQF0}) has weight zero; in particular, the corresponding equation 
for $F(\yc)$ has no $\ell$ dependence. 

Now the idea is to consider a UV ansatz for $\zeta(\rad,\theta)$ of the type,
\beq
\zeta(\rad,\theta)= \frac{\eLL^2}{R^4}\,\Func\left(\yc,\theta \right)\ . \label{UVANSATZ}
\eeq  
As expected, by plugging (\ref{UVANSATZ}) into the equation of motion we obtain an equation for 
$\Func(\yc,\theta)$ which depends only on the dimensionless parameter 
$\varepsilon = \frac{\Delta}{R^2}$ for D3 and D1-D5 branes with $\Delta\equiv z_1\eLL $, and
$\varepsilon = \frac{\Delta}{R^{3/2}}$ for M2 branes with $\Delta=z_1\sqrt{\eLL}$. The limit 
$\varepsilon \to0$ is well defined and gives back (\ref{EQF0}). 
Around it we can solve the equation for $\Func(\yc,\theta)$ in perturbation theory. Schematically, 
our problem becomes 
\bea
\mathrm{Eq}\big[\Func_p(\yc,\theta), \Delta\big]&=&0\, , \nn\\ 
\Func_p(\yc,\theta)&=&(1-\yc^2)+\sum_{k=1}^{\infty} \Delta^k f^{(k)}_p(\yc,\theta)\, . 
\label{SERIESUV}
\eea
In (\ref{SERIESUV}) we restored the label $p$ to stress that the perturbative solution depends on 
the number of dimensions. The functions $f^{(k)}_p$ capture the two-center deformation of the UV 
$AdS$ solution. Solving for $f_p^{(k)}$ still requires finding the solution of a set of PDEs. 
However, this problem is tractable and analytic solutions can be obtained.

\subsubsection*{Perturbative equations}

For D3 and D1-D5 branes it is possible to write down simple explicit formulae. 
Results for the M2 branes are more involved due to the fact that the UV $AdS$ comes in 
horospherical coordinates. However, the algorithm to find the perturbative solution is valid for 
generic $p$.
 
For $p=1,3$, the functions $f_p^{(k)}$ solve a PDE of the form,
\beq\label{EQPDEUV}
\partial_\yc^2 f_p^{(k)} +\frac{p-1}{\yc(1-\yc^2)}\partial_\yc f_p^{(k)}
+\frac{1}{(1-\yc^2)^2}\left(\partial_\theta^2 f^{(k)}_p +(p+1)\cot\theta\,\partial_\theta f^{(k)}_p \right)
=\mathcal{F}^{(k)}(\yc,\cos\theta)
\eeq
where $\mathcal{F}^{(k)}$ are forcing terms whose explicit $\theta$ dependence is inherited from 
$\mathcal{H}=\mathcal{H}(\zeta,\cos\theta)$. At fixed $k$, the forcing term $\mathcal{F}^{(k)}$ is 
determined by the lower order solutions $f_p^{(m)}$ for $m<k$. We find the first non-trivial 
$\mathcal{F}^{(k)}$, and solve for $f_p^{(k)}$. Then we proceed to compute $\mathcal{F}^{(k+1)}$, 
solve for $f_p^{(k+1)}$, and continue by iteration. An important observation is that upon the change 
of variable $v=\cos\theta$, the forcing terms $\mathcal{F}^{(k)}$ become polynomials in $v$ with 
$\yc$-dependent coefficients. Therefore, the ansatz
\beq\label{ANSATZPERT}
f^{(k)}_p= g^{(k,k)}_p(\yc) v^k + g^{(k,k-1)}_p(\yc)v^{k-1}+ \ldots + g^{(k,0)}_p
~,
\eeq 
which is compatible with the boundary conditions $\partial_{\theta} f^{(k)}_p=0$ at $\theta=0,\pi$,
solves the $\theta$ dependence in (\ref{EQPDEUV}). The set of functions $\{v^m\}_{m=0}^{\infty}$ 
is just a rewriting of the standard Fourier basis in a way that is compatible with our boundary 
conditions. For any $f^{(k)}_p$ of the form (\ref{ANSATZPERT}), the PDE  (\ref{EQPDEUV}) 
generates a set of $k$ ODEs for the functions $\{ g^{(k,n)}_p \}_{n=0}^{k}$. The boundary 
conditions that uniquely specify the solution of each $g^{(k,n)}_p(\yc)$ are  
\beq
g^{(k,n)}_p(\yc=1)=0\, ,\qquad \partial_\yc g^{(k,n)}_p(\yc=0)=0\ .
\eeq  
The use of the coordinate $\zeta$ makes manifest the fact that in order to have a perturbative solution 
which is consistent with the UV $AdS$ asymptotics, the functions $g^{(k,n)}_p$ have to vanish like 
$(1-\yc^2)^\alpha$ with $\alpha\ge1$. When $\alpha>1$, corrections will be sub-leading at the boundary. 

The equations for the functions $g^{(k,n)}_p$ are linear ODEs with forcing terms induced by 
$\mathcal{F}^{(k)}$. The highest mode $g^{(k,k)}_p$ has no forcing term. At fixed $n<k$, the 
equation for $g^{(k,n)}_p$ has forcing terms induced by the functions $g^{(k,m)}_p$ with 
$n<m\le k$. Starting from $n=k$ and solving for $g^{(k)}_p$ it is possible to generate the 
forcing term for $g^{(k-1)}_p$ and solve its equation. At the next step we generate the 
forcing terms for $g^{(k-2)}_p$ and solve its equation. Repeating this algorithm it is 
possible to calculate the full tower of $\{ g^{(k,n)}_p \}_{n=0}^{k}$ modes. 

We conclude this subsection with one relevant comment: there is no $f_p^{(1)}$ contribution to the 
perturbative solution. This statement follows from: $1)$ the fact that the equation of motion 
depends just on $\mathcal{H}^2$, $2)$ the expansion of $\mathcal{H}$ in terms of $\Delta$ 
is given by the UV expansion of the harmonic functions (\ref{MALDAD1D5}) and 
(\ref{MALDAD3M2}), and $3)$ in the latter, the contribution at order $\Delta$ is proportional to the center of mass condition and therefore vanishes.

\subsection{Two-centered D3 geometries}

We are now in position to carry out the perturbative calculation in the two-centered D3 brane solution
more explicitly. The analytic result for $\Func_3(\yc,\theta)$ can be written in a compact form by 
defining the variable $\X=(1-\yc^2)$. The first non-trivial corrections to $\surfRT$ are
\bea
\Func_3(\yc,\theta)- \X &=& - \frac{2}{3}\frac{N_1}{N_2} \left( 6\cos^2\theta -1 \right) \X^2 
\left( \frac{\Delta}{R_{UV}^2} \right)^2    
\nn\\
& &  +\frac{ N_1-N_2}{N_2}\frac{N_1}{N_2} 
\left( 8\cos^3\theta -3\cos\theta \right) \X^{5/2} \left( \frac{\Delta}{R_{UV}^2} \right)^3     
\nn\\
 & &  +\left( g^{(4,4)}\cos^4\theta+ g^{(4,2)}\cos^2\theta+ g^{(4,0)}\right) \left(\frac{\Delta}{R^2_{UV}}\right)^4 
 +\, \ldots \,  \label{SOLUVD3}
\eea
where 
\bea
g^{(4,4)}&=& -\frac{16N_1}{N_2} \frac{N_1^2-3N_1 N_2 +N_2^2}{N_2^2} \X^3\ ,\\
g^{(4,2)}&=& \frac{16N_1}{N_2} \frac{9N_1^2-17N_1 N_2 +9N_2^2}{15 N_2^2} \X^3\ ,\\
g^{(4,0)}&=&\frac{4}{9} \frac{N_1^2}{N_2^2} X^2
- \frac{N_1}{N_2} \frac{27N_1^2-71N_1N_2+27N_2^2}{45N_2^2}X^3
\ .
\eea
It is intriguing that $f^{(2)}(X,\theta)$ and $f^{(3)}(X,\theta)$ are separable, whereas $f^{(4)}(X,\theta)$ is not. 
In general, higher modes $f^{(k)}$ with $k\ge4$ are also not separable. We will come back to this aspect of 
the solution later on. Finally, we could have guessed from the beginning that when $N_1=N_2$ a symmetry 
argument implies that $f^{(k)}$ with $k$ odd will be vanishing.

Plugging the solution (\ref{SOLUVD3}) into the HEE functional given by equations 
(\ref{ACTIONUV})-(\ref{UVLAGR}) we obtain
\bea\label{EEUVD3}
S_3&=&\frac{1}{4G_N^{(5)}}\, \left(4\pi R^3_{UV}\right)\, \left( I_3(\eLL) - \frac{4}{9} \frac{N_1^2}{N_2^2}\left(\frac{\Delta}{R^2_{UV}}\right)^4\, +\, \ldots \right)\ ,\\
I_3(\eLL)&=&  \int_0^1 d\yc\, \frac{\yc^2}{(1-\yc^2)^2}=\ \int_{a/\ell}^1 d\yh\, \frac{\sqrt{1-\yh^2}}{\yh^3}.
\eea
In (\ref{EEUVD3}) we used the relation
\beq
G_N^{(5)}=\frac{G_N^{(10)}}{\pi^3 R_{UV}^5} \ .
\eeq
The integral $I_3(\eLL)$ is the $AdS$ \RT result \cite{Ryu:2006bv} with $a/\ell$ their UV cutoff. 
Surprisingly, even though the profile of the surface gets corrections at order $\Delta^2$ and $\Delta^3$, 
the first non-vanishing contribution to the entanglement entropy comes at fourth order. 
Higher order correction are also non-trivial but their expression is too cumbersome and not 
sufficiently illuminating to repeat here. In agreement with the expectation that the renormalized 
entanglement entropy decreases along the RG flow, the first non-trivial correction to $I_3(\eLL)$ 
in (\ref{EEUVD3}) comes with a negative sign.   

Geometrically, the reason why there are nor $\Delta^2$ neither $\Delta^3$ corrections to the HEE can 
be seen as follows. We first observe that 
\beq\label{HARMONICSD3}
\left( 6\cos^2\theta -1 \right)\, \propto\, Y^{(5)}_{\,\vec{0},2}(\theta)\,,\qquad 
\left( 8\cos^3\theta -3\cos\theta \right)\, \propto\, Y^{(5)}_{\,\vec{0},3}(\theta)\, ,
\eeq
where $Y^{(5)}_{\vec{0},l}$ are the $S^4$-invariant $5$-dimensional spherical harmonics. Then, 
we notice that the expression of the integrand of $S_3$, at order $\Delta^2$ or $\Delta^3$, takes the 
form of a scalar product\footnote{The measure in the scalar product is $\sqrt{g_{ij}}$ on the $S^5$.} 
between the harmonics (\ref{HARMONICSD3}) and the identity. In particular we find, 
\beq
S_3 \sim  I_3(\eLL)  + \langle 1\, | Y_{\vec{0},2} \rangle \left(\frac{\Delta}{R^2_{UV}}\right)^2 +  
\langle 1\, | Y_{\vec{0},3} \rangle 
\int_0^1 d\yc\, \yc^2\sqrt{1-\yc^2}  
\left(\frac{\Delta}{R^2_{UV}}\right)^3
+ \ldots  \label{EESCALARPROD_D3}
\eeq 
The result (\ref{EEUVD3}) follows from the orthonormality condition 
$\langle Y_m|Y_n\rangle=\delta_{mn}$. We would like to stress that the decomposition 
(\ref{EESCALARPROD_D3}) is not immediately obvious, and it comes out from the interplay 
between the UV expansion of the metric and the form of the solution. 

The use of the scalar product between harmonics may be a useful way of packaging the expansion 
of the HEE. It also suggests that in order to have non-vanishing corrections, we should find at least 
terms of the type $Y^2$. The only way to generate such contributions is through the non-linearity 
of the background metric, and indeed multi-centered geometries are non-linear solutions. 

As we briefly reviewed in section \ref{multicenterUV}
the field theory description of the two-centered D3 solution is well understood at the UV. 
By splitting the stack of coincident branes along the $z$ direction, we give an expectation 
value to one of the real adjoint scalar fields of $\mathcal{N}=4$ SYM. Therefore, the 
1-point function of the gauge invariant chiral operators $\mathcal{O}^{(n)}$, defined in 
(\ref{CHIRAL}), will be non-trivial. Given the relation between these operators and the 
harmonics of $S^5$, it is possible to show that the AdS/CFT correspondence correctly 
reproduces the 1-point function of the operators $\mathcal{O}^{(n)}$ 
unambiguously \cite{Kraus:1998hv}. This fact invites us to think of the 
result (\ref{EESCALARPROD_D3}) as the statement that at small distances corrections to the
entanglement entropy associated to $\mathcal{O}^{(2)}$ and $\mathcal{O}^{(3)}$ vanish.
It would be interesting to examine this possibility directly in field theory.

\subsection{Two-centered D1-D5 geometries}

In this section we repeat the perturbative computation of the entanglement entropy in two-centered 
D1-D5 geometries producing a prediction for the corresponding two-dimensional conformal field theories.

\begin{figure}[t]
\centering
\rule{.01cm}{.1pt}
\includegraphics[scale=.44]{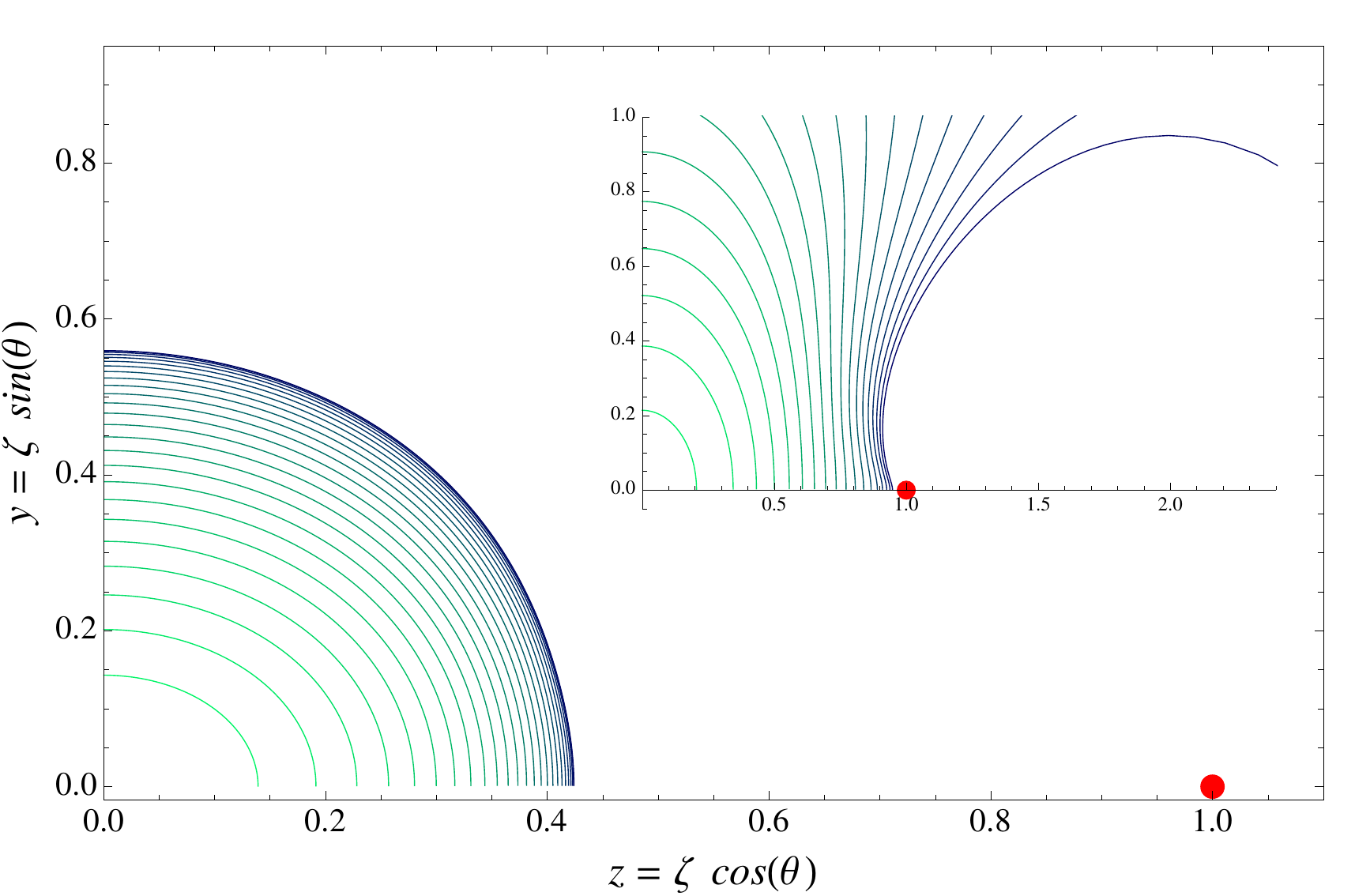}
\rule{.01cm}{0pt}
\includegraphics[scale=.44]{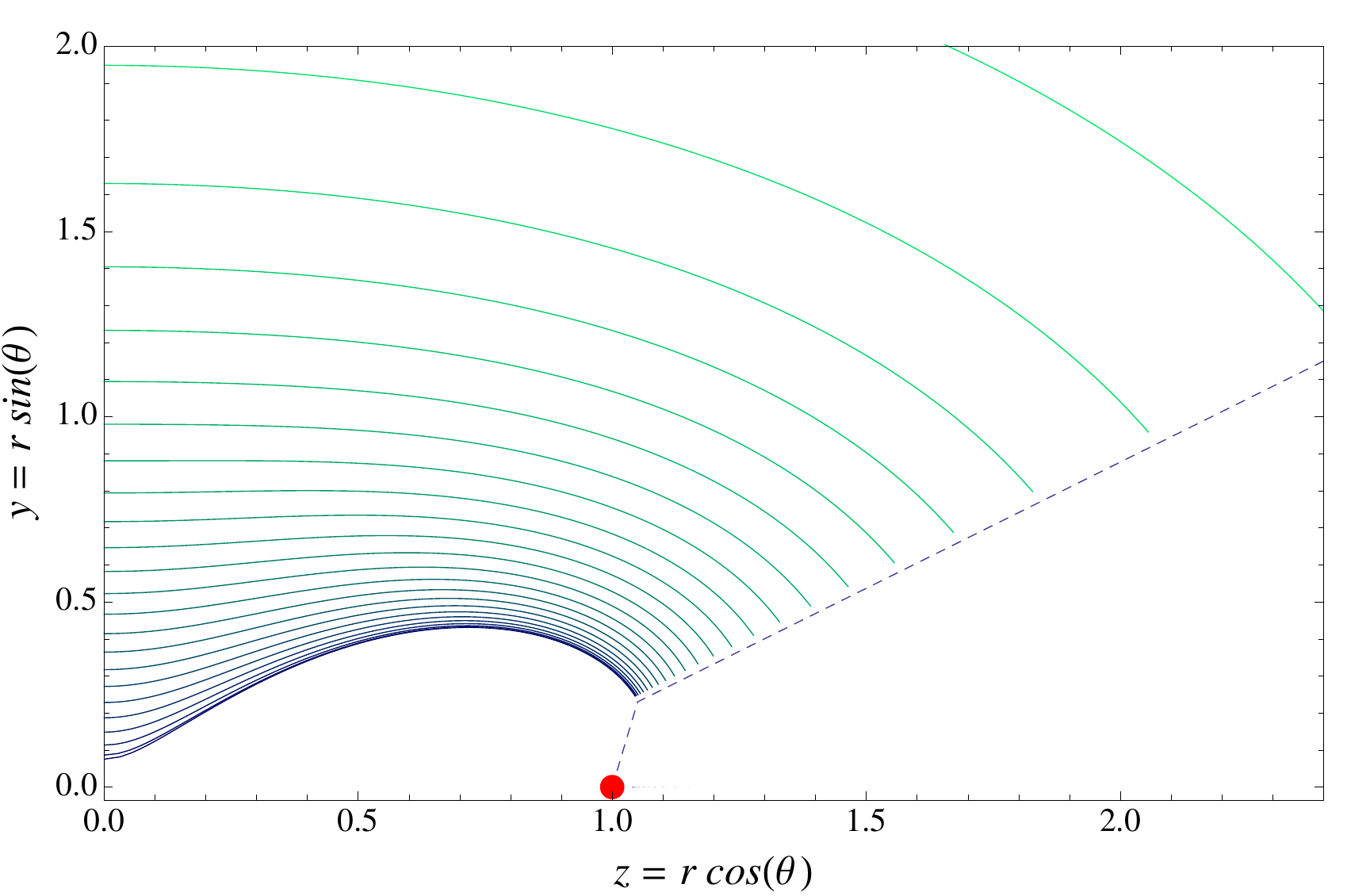}
\caption{ {\it\small
In the upper figure we plot the transverse scalar function 
$\Func_{1\cup 5}\,$ in the $(z,y)$ plane for $\Delta=2$ and $z_1=-z_2=1$. We are using the  
polar coordinates $z=\zeta \cos \theta$ and $y=\zeta \sin\theta$. The location of the branes is 
indicated by a red dot. The UV boundary is at the origin. The foliation corresponds to equally 
spaced intervals in $(0,\ell)$, and is approximately made by circles. For this value of $\Delta$ 
the equation of motion is satisfied with a minimum accuracy of $\,10^{-5}$. In the inset we 
show an extrapolation to a value of $\Delta$ which comes closer to the formation of the separatrix.
In the lower figure we use coordinates $z=r \cos \theta$ and $y=r \sin\theta$ with $r=1/\sqrt{\zeta}$. 
In this specific plot the function $\Func_{1\cup 5}\,$ is extrapolated to $\Delta=2.755$. The r.h.s.\ 
part of the plot, where the solution is less reliable, has been excised. The qualitative features of 
this solution agree with the features anticipated in the general discussion in Section 
\ref{HEEMULTICENTER}. We see how the surface deforms around the centers and how 
the turning point of the surface approaches a separatrix.
}}
\label{FIGURE2_D1D5}
\end{figure}

Keeping the notation $\X=(1-\yc^2)$, the analytic form of $\Func_{1\cup 5}$ up to fourth order is 
\bea
\Func_{1\cup 5}(\yc,\theta)- \X &=& - \frac{2}{3}\frac{Q_1}{Q_2} \left( 4\cos^2\theta -1 \right) \X^2 
\left( \frac{\Delta}{R_{UV}^2} \right)^2    \nn\\
& &  +\frac{ Q_1-Q_2}{Q_2}\frac{Q_1}{Q_2} \left( 4\cos^3\theta -4\cos\theta\right) \X^{5/2} 
\left( \frac{\Delta}{R_{UV}^2} \right)^3     \nn\\
& &  +\left( g^{(4,4)}\cos^4\theta+ g^{(4,2)}\cos^2\theta+ g^{(4,0)}\right) 
\left(\frac{\Delta}{R^2_{UV}}\right)^4 \, ,
\eea
where 
\bea
g^{(4,4)}&=& -\frac{16Q_1}{Q_2} \frac{18Q_1^2-49Q_1 Q_2 +18Q_2^2}{45Q_2^2} \X^3~,
\\
g^{(4,2)}&=& -\frac{16Q_1^2}{135Q_2^2}\X^2+ \frac{24}{5}\frac{Q_1}{Q_2}\frac{(Q_1-Q_2)^2}{Q_2^2} \X^3~,
\\
g^{(4,0)}&=&\frac{8}{15} \frac{Q_1^2}{Q_2^2} X +\frac{22}{135} \frac{Q_1^2}{Q_2^2}X^2
- \frac{Q_1}{Q_2} \frac{18Q_1^2-53Q_1Q_2+18Q_2^2}{45Q_2^2}X^3
~.
\eea
As we found in the case of the D3 brane solution, the $\yc$ and $\theta$ dependence of $f_1^{(2)}$ 
and $f^{(3)}_1$ factorizes and we can write 
\beq\label{D1D5HARMONICS}
f_1^{(2)}\, \propto\, Y^{(3)}_{\vec{0},2} \X^2\,,\qquad\qquad   f_1^{(3)}\, \propto\, 
Y^{(3)}_{\vec{0},3} \X^{5/2}\,,
\eeq
where $Y^{(3)}_{\vec{0},l}$ are harmonics of $S^3$ symmetric with respect to the $\vec{\Omega}$ 
angles. 

From the series expansion of $\Func_{1\cup5}$, we obtain a series expansion for the HEE. At lower 
orders we find
\bea\label{EEUVD1D5}
S_1&=&\frac{R_{UV}}{4G_N^{(3)}}\, \left( I_1(\eLL) 
- \frac{1}{20} \frac{Q_1^2}{Q_2^2}\left(\frac{\Delta}{R^2_{UV}}\right)^4\, +\ldots \right) 
\\
I_1(\eLL)&=&\int_0^1 d\yc\, \frac{2}{1-\yc^2}=\, 2\, \int_{2a/\ell}^{\frac{\pi}{2}}  \frac{ds}{\sin s}\,
\eea
where we used the relation
\beq
G_N^{(3)}=\frac{G_N^{(6)}}{2\pi^2 R_{UV}^3}\ .
\eeq
The integral $I_1(\eLL)$ is the \RT result \cite{Ryu:2006bv}, and the first non-trivial correction 
comes at fourth order, as in the case of the D3 brane system. Along the lines of 
(\ref{EESCALARPROD_D3}), we find that the vanishing of $\Delta^2$ and $\Delta^3$ 
corrections can be interpreted as the vanishing of the scalar product between different harmonics.

In addition, we computed $\Func_{1\cup5}$ for a D1-D5 system with $Q_1=Q_2$ up to 
order $\Delta^{18}$, and studied the convergence of the series. We checked explicitly 
that at orders $k>3$, separation of variables does not occur for any $f^{(k)}$. 
Because our perturbative expansion makes use of a spectral decomposition, it works 
quite well in a certain range of $\Delta$. An example is given in Figure \ref{FIGURE2_D1D5},
where we observe that the qualitative features of the solution agree with the features anticipated in 
section \ref{HEEMULTICENTER}.
 
\begin{figure}[t]
\centering
\includegraphics[scale=.45]{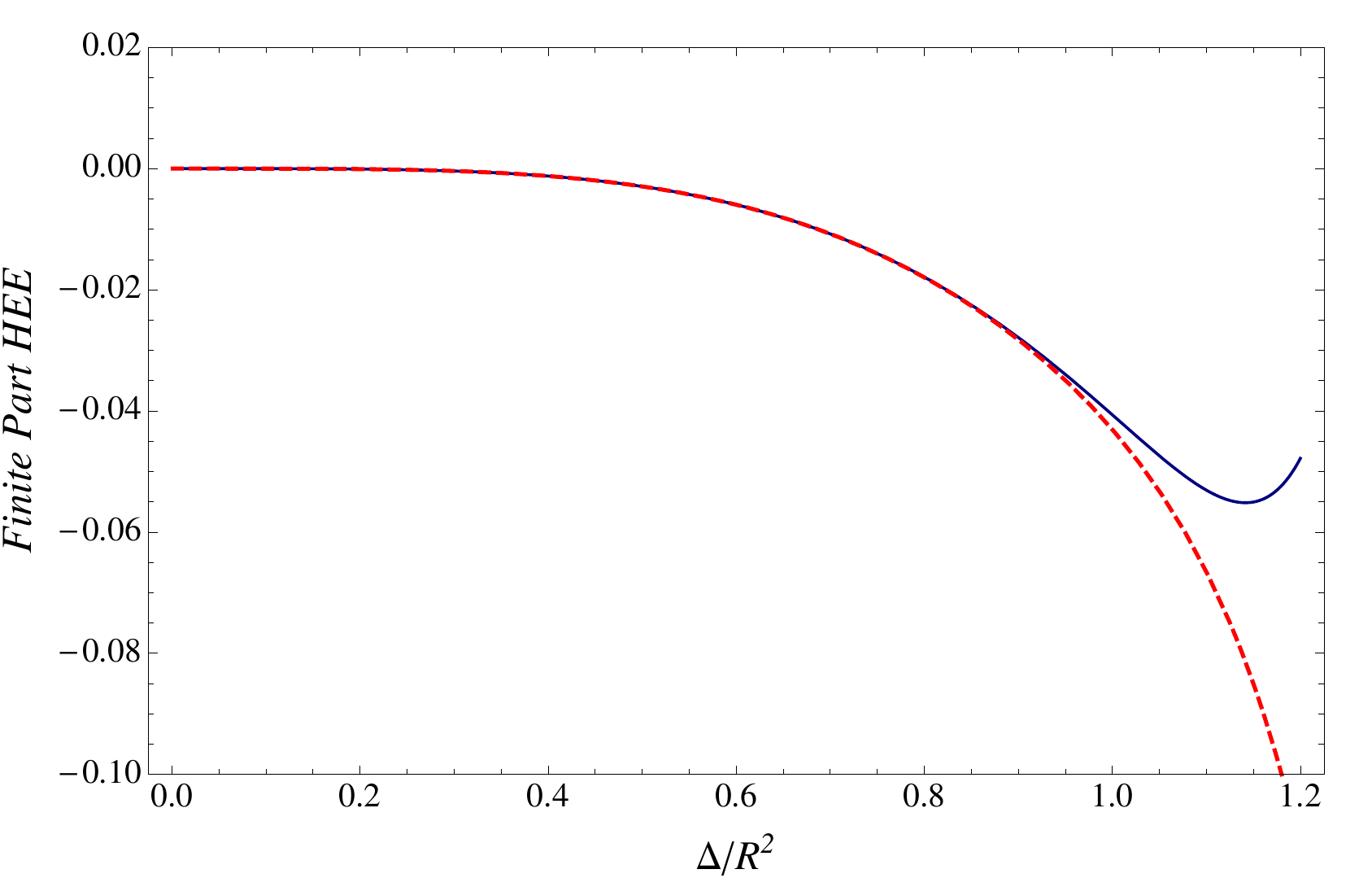}
\caption{ {\it\small
We plot the finite part of the HEE defined by subtracting $I_1(\ell)$. In units of $R_{UV}/4G_N^{(3)}$, 
the blue (red) curve represents the series expansion up to order $\Delta^{18}$ $\left(\Delta^{16}\right)$. 
The embedding function $\zeta$ and the HEE have different sensibility with respect to $\Delta$. 
} }
\label{FIGURE2_D1D5C}
\end{figure}

After subtracting $I_1(\eLL)$ the HEE of the D1-D5 system is expressed as a series expansion in 
$\Delta$ with coefficients that can be determined analytically. The resulting series is alternating. For 
example, the coefficient of $\Delta^{2k}$ for $k=2,\ldots,7$ are,
\beq
\Big\{ \textstyle{ -\frac{1}{20}\, ,~ \frac{8}{567} \, ,~ -\frac{1.567}{170.100}\, ,~ \frac{40.729}{7.016.625}\, , -\frac{101.669.532}{23.508.883.125}\,,~ \frac{30.609.041.679}{9.050.920.003.125}\,,~\ldots }\Big\}\ .
\eeq
The corresponding curve is also plotted in Figure \ref{FIGURE2_D1D5C}.

%
\subsection{Two-centered M2 geometries}

We conclude this section by analyzing the perturbative solution of $\Func_2$ for 
two-centered M2 brane geometries. The notation is unchanged, $\X=(1-\yc^2)$. 
The leading contributions to the embedding function are given by
\bea
\Func_2(\yc,\theta)-\X &=& -3\frac{M_1}{M_2} \left( 8\cos^2\theta-1 \right) X^{3/2} 
\left( \frac{\Delta}{R_{UV}^{3/2} } \right)^2\\
& & +\,  64\,\sqrt{2}\,\frac{M_1-M_2}{M_2}\frac{M_1}{M_2}
\left( \cos^3\theta - \frac{3}{10} \cos\theta \right) X^{7/4}\left( \frac{\Delta}{R_{UV}^{3/2} } \right)^3 \\
& &  +\left( g^{(4,4)}\cos^4\theta+ g^{(4,2)}\cos^2\theta+ g^{(4,0)}\right) 
\left(\frac{\Delta}{R^{3/2}_{UV}}\right)^4 \, +\ldots 
\eea
where $R_{UV}$ is the radius of the UV $AdS$ and,
\bea
g^{(4,4)}&=& -\frac{32 M_1}{M_2} \frac{10M_1^2-37 M_1 M_2 +10M_2^2}{M_2^2} \X^2\\
g^{(4,2)}&=& \frac{20 M_1}{M_2}\frac{(8M_1^2-17 M_1 M_2 +8 M_2^2)}{M_2^2} \X^2\\
g^{(4,0)}&=& - \frac{21 M_1^2}{M_2^2} \left( \sqrt{X} - \log\left(1+\sqrt{X}\right)-\frac{X}{2}\right) 
- \frac{M_1}{M_2} \frac{32M_1^2-89 M_1M_2+32M_2^2}{4M_2^2}X^2
~. \nn
\\ 
\eea
Certain features of $\Func_2$ are similar to the previous cases. In particular, 
we find for any $p$ that the corrections $f^{(2)}_p$ and $f^{(3)}_p$ are solved 
by separation of variables. The origin of this feature is unclear. It is possible that supersymmetry 
is related to this effect (recall that we are studying BPS configurations).

The HEE  expanded at lower orders is
\bea\label{EEUVM2}
S_2&=&\frac{1}{4G_N^{(4)}}\,  (2\pi R_{UV}^2)\left( I_2(\eLL) 
- \frac{35}{4} \frac{M_1^2}{M_2^2}\left(\frac{\Delta}{R^{3/2}_{UV}}\right)^4\, +\ldots \right) \\
I_2(\eLL)&=&\int_0^1 d\yc\, \frac{\yc}{(1-\yc^2)^{3/2}}=\, \int_{a/\ell}^{1}  \frac{d\yh}{\yh^2}\,
\eea
where the lower-dimensional Newton constant is,
\beq
G_N^{(4)}=\frac{G_N^{(11)}}{\pi^4/3\, R_{S^7}^7}
~ .
\eeq
In defining $G_N^{(4)}$ we made use of the relation $R_{S^7}=2R_{UV}$. 
The expression (\ref{EEUVM2}) again shows that the first non-trivial correction to the \RT result 
$I_2(\eLL)$ \cite{Ryu:2006bv} comes at fourth order.

\section{IR expansion of the entanglement entropy}
\label{HEE2ir}

As we increase the radius $\ell$ of the entangling surface, the bulk minimal surface $\surf$ 
starts to probe the interior of the $D$-dimensional bulk geometry. For a given $\ell_c$, the 
surface hits the branching point, and for $\ell\ge\ell_c$ the topology of $\surf$ is that 
of a pant with two legs. 
Geometrically, for $\ell\ge\ell_c$, 
the surface is ``attracted" towards the position of the branes. 
The qualitative picture to keep in mind is given by Figure \ref{FIGURA}.

Target space coordinates adapted to the center-of-mass become problematic if we want 
to describe $\surf$ in Phase \DUE. Below the separatrix $r(\rad,\theta)$ is double-valued 
as a function of $\theta$ for fixed $\rad$ in a neighborhood of $\rad=0$. To overcome this 
problem we will use a different system of coordinates. This is also motivated by the following 
field theory observation. The end-point of the RG flow is a collection of decoupled theories, 
therefore the leading contribution to the entanglement entropy in the deep IR has to be the 
sum of the entanglement entropies of each individual throat. This expectation implies that as 
$\ell\to\infty$, the contribution to the area of $\gamma$ coming from the patch 
outside the separatrix has to become subleading. We will see how the new coordinate 
system clarifies the role of the separatrix as we take the deep IR limit.

\subsection{Adapted coordinates}\label{SEC_COORD}

We first focus on two-centered geometries with $\mathbb{Z}_2$ symmetry, namely 
$z_1=-z_2\equiv \bz$. The 
change of coordinates relevant for this case is constructed as follows. Starting from the 
hyper-cylindrical coordinates $(z,y)$ we introduce 
\begin{itemize}
\item[1)] polar coordinates $z=r\cos\theta$ and $y=r \sin\theta$, 
\item[2)] we define the $(u,v)$ variables by means of the relation, 
\beq\label{OBSTACLEMAP}
u+iv=\left(\sqrt{ \left(z+ i y\right)^2 -\bz^2 }\right)^2 ,
\eeq
which is equivalent to
\beq\label{OBSTACLEMAPINV}
r^2=\sqrt{\left( u+\bz^2\right)^2+v^2}\, ,\qquad \theta=\frac{1}{2}\arctan \left(\frac{v}{u+\bz^2}\right),
\eeq
and finally, 
\item[3)] we consider polar coordinates $u=\eta\cos\psi$ and $v=\eta\sin\psi$. 
\end{itemize}

\begin{figure}[t]
\centering
\includegraphics[scale=.5]{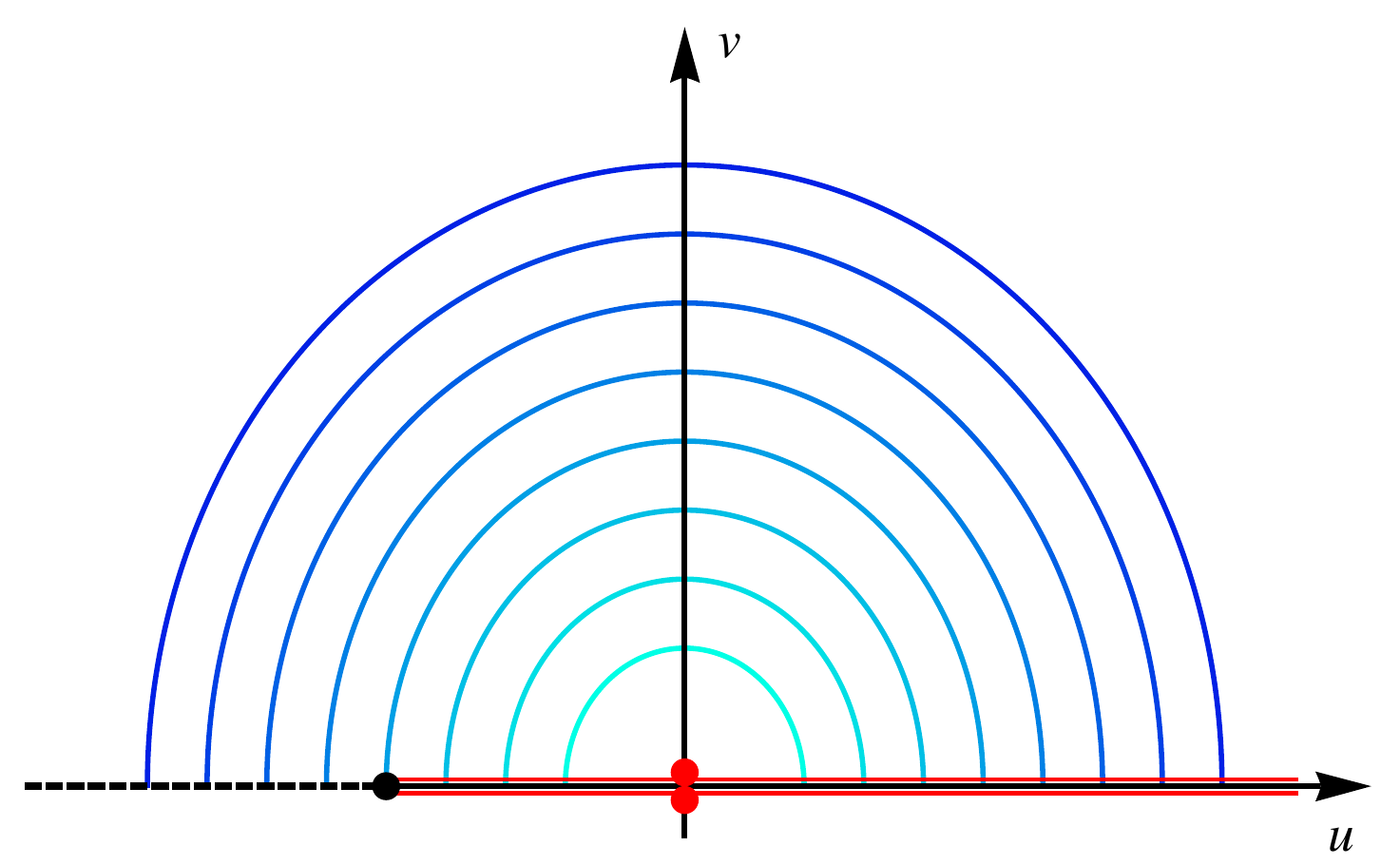}
\rule{.25cm}{0pt}
\includegraphics[scale=.5]{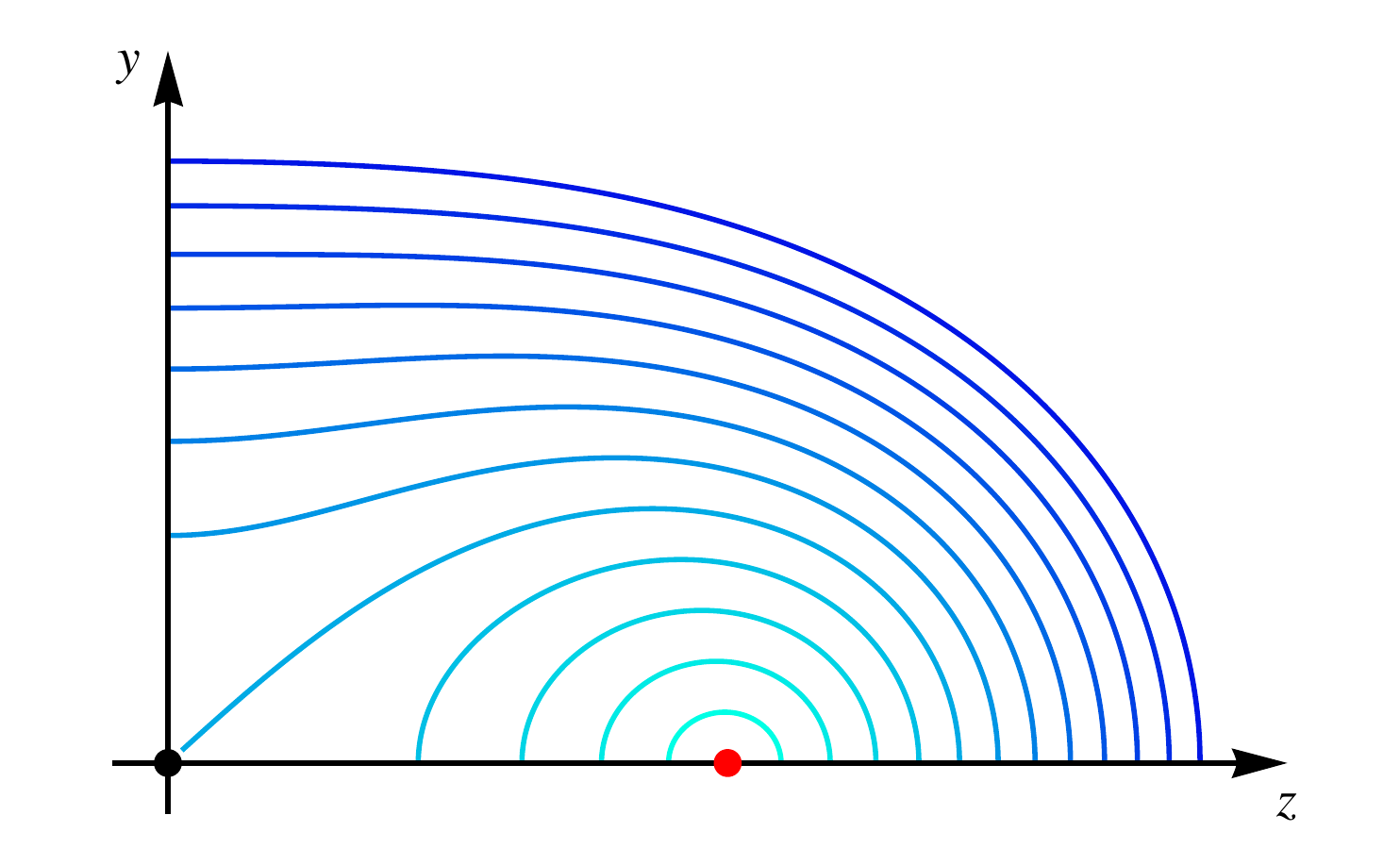}
\caption{ {\it\small Circles in the $(u,v)$ plane (l.h.s.\ picture) are mapped to closed curves in the $(z,y)$ 
plane (r.h.s. picture). The red dots indicate the position of the branes. The black dot in the $(u,v)$ plane is 
mapped to the origin in the $(z,y)$ plane. 
}}
\label{FIGURE3}
\end{figure}

The geometry in the $(u,v)$ plane is such that the two stacks of branes are both located at the origin, 
$u=0$ and $v=0^{\pm}$, one in the upper half plane and the other in the lower half plane. 
The $\mathbb{Z}_2$ symmetry has become a reflection symmetry between these two planes. 
From the relation (\ref{OBSTACLEMAP}), it is simple to see that the interval $\{|z|\le \bz,y=0\}$ has 
been mapped to $\{-\bz^2<u<0,\, v=0^{\pm}\}$, whereas the $y$-axis $\{y>0,z=0\}$ and semi-infinite 
lines $\{ |z|\ge\bz, y=0\}$ have been mapped to $\{ u < -\bz^2,v=0\}$\footnote{The determination of 
the $\arctan$ in (\ref{OBSTACLEMAPINV}) has to be chosen correctly when $u>-\bz^2$ and  $u<-\bz^2$.} 
and $\{u>0,v=0^{\pm}\}$, respectively. 
See Figure \ref{FIGURE3} for an illustration. Geodesics can cross the line $\{ u<-\bz^2,v=0\}$, and 
go from the upper to the lower half plane. The lines $\{u>-\bz^2,v=0^{\pm}\}$, instead, are a boundary.
The change of variables (\ref{OBSTACLEMAP}) is  borrowed from 2d complex analysis \cite{Lang}.

\vspace{0.3cm}
\paragraph{UV and IR limits.}
As an example, the two-centered D1-D5 metric with $Q_1=Q_2\equiv Q$ has the following translation 
in the new coordinates
\bea
H_{1\cup5}\left( dz^2+ dy^2\right)\,&\to&\,  \frac{R^2_{UV}}{4\eta^2} 
\left( 1+\frac{\bz^2}{\sqrt{\bz^4+\eta^2+2\bz^2\eta\cos\psi}}\right)\left( d\eta^2+\eta^2 d\psi^2 \right) \label{D1D5EX1}
~,\\
H_{1\cup 5}^{-1}\, dx^2\, &\to &\, \frac{\,\eta^2}{R^2_{UV}}\frac{dx^2}{\left(\bz^2
+\sqrt{\bz^4+\eta^2+2\bz^2\eta\cos\psi}\right)}\ . \rule{.0pt}{.8cm}    \label{D1D5EX2}
\eea
Formulas \eqref{D1D5EX1}-\eqref{D1D5EX2} are useful as concrete reference for the subsequent calculations. 
However, the discussion that follows is general, and it holds for any $p$, i.e. for $D3$ and $M2$ branes as well.

Describing the Coulomb branch in this coordinate system is advantageous because 
the UV and the IR limits of the geometry can be formally explored by sending 
$\eta\to\infty$ and  $\eta\to0$, as in the case of coincident branes. 
In the limit $\eta\to\infty$ we recover the UV $AdS\times S$ geometry with radius $R_{UV}$,
\bea \label{UVOBSTACLED1D5}
ds^2_{UV}&=&\ \frac{\eta}{R^2_{UV}} dx^2 + \frac{R^2_{UV}}{4\eta^2}d\eta^2
+ R_{UV}^2\left( \frac{d\psi^2}{4}
+ \sin^2\frac{\psi}{2}\, d\vec{\Omega}^2\right) 
\\
&\hookrightarrow &\rule{.3cm}{.0pt}\, R^2_{UV} 
\left( \rho^2 dx^2+\frac{d\rho^2}{\rho^2} + \frac{d\psi^2}{4} +  \sin^2\frac{\psi}{2}\, d\vec{\Omega}^2 \right)				\quad \mathrm{\ with}\quad\eta=R^4_{UV}\rho^2
~.
\eea  
In the $\eta\to0$ limit we obtain the metric  
\beq\label{IROBSTACLED1D5}
ds^2_{IR}=\frac{1}{4R^2_{IR}}\frac{\eta^2}{\bz^2}dx^2+R^2_{IR}\frac{d\eta^2}{\eta^2}
+R^2_{IR}\left( d\psi^2+\sin^2\psi d\vec{\Omega}^2 \right)
~.
\eeq

It is important to point out two facts about (\ref{UVOBSTACLED1D5}) and (\ref{IROBSTACLED1D5}). 
The first is that the metric (\ref{IROBSTACLED1D5}) is described by a radial coordinate 
which is essentially a double covering of the UV $AdS$. The second is that the metric 
(\ref{IROBSTACLED1D5}) still depends on $\bz$ and therefore we need to properly define 
the IR limit. In fact, from the field theory side we know that in the limit $\bz\to\infty$ the theory 
is decoupled at all energy scales and consists of two independent SCFTs. However, taking 
the limit $\bz\to\infty$ in (\ref{IROBSTACLED1D5}) does not return an $AdS$ solution. This issue 
is simply solved by defining
\beq
\etair=\eta/\bz\ .
\eeq 
The correct IR limit is then obtained by keeping the variable $\etair$ fixed, while taking 
the limit $\bz\to \infty$. This prescription gives the IR $AdS$ as the zeroth order metric of a $1/\bz$ 
expansion,
\beq\label{CORRECTIR}
ds^2=\frac{\etair^2}{4R^2_{IR}}dx^2+R^2_{IR}\frac{d\etair^2}{\etair^2}
+R^2_{IR}\left( d\psi^2+\sin^2\psi d\vec{\Omega}^2 \right)+\mathcal{O}\left(\frac{1}{\bz}\right)\, .
\eeq
All corrections vanish in the limit $\bz\to \infty$ and we recover the expected decoupling of 
the full geometry. 

At this point, it is also useful to write down the expression for the $\surfRT$ 
surface embedded in the metric (\ref{CORRECTIR}). 
The equation of motion and the solution of $\eta(\rad)$ are,  
\bea
\etair''+\left[ \frac{p-1}{x}-({p+2})\frac{\etair'}{\etair}
+4R_{IR}^4\frac{p-1}{x}\frac{\etair'^2}{\etair^4}\right]\etair'-\frac{p}{4R^4_{IR}}\etair^3=0
\label{IREQUATIONZERO}\\
\etair(\yc)= \frac{1}{\ell}\frac{2R_{IR}^2}{\sqrt{1-\yc^2} }\qquad \mathrm{with}\qquad \yc
=\rad/\ell
~. 
\label{IRESOLUNO}
\eea
On the other hand, the embedding function for $\surfRT$ in the UV AdS  is easily obtained 
from the solution (\ref{ADS_SOL}) by noticing that  (\ref{UVOBSTACLED1D5}) gives the 
$AdS$ metric (\ref{UVADSUV}) after the change of variables $\eta=1/\zeta$. We thus find 
the relation
\beq\label{SQUAREPROP}
\eta_{UV}= \frac{R^4_{UV}}{\ell^2 \left(1-\yc^2\right)}\, ,\qquad \etair
=\frac{2 R_{IR}^2}{R_{UV}^2} \sqrt{\eta_{UV}}
~.
\eeq
The property (\ref{SQUAREPROP}) fits naturally with the observation that 
(\ref{IROBSTACLED1D5}) is a double covering of (\ref{UVOBSTACLED1D5}).

\subsection{Details of the IR expansion}

The original embedding function $r(\rad,\theta)$ described in Section \ref{HEEMULTICENTER} 
becomes in the new coordinates $\eta=\eta(\rad,\psi)$. This function is always single-valued as 
a function of $\psi$. Exploiting the symmetry of the $\mathbb{Z}_2$ symmetric solution
we can restrict $\psi\in(0,\pi]$ and impose appropriate boundary conditions at $\psi=\pi$. 
 
The minimal surface is governed by the Euler-Lagrange equations of a Lagrangian with the 
structure of (\ref{UVLAGR}). For quick reference we repeat here the specifics of the D1-D5 case,
\bea
\mathcal{L}_{1\cup5}&=&\, \mathcal{K}[\psi,\eta]\,\mathcal{H}[\psi,\eta]\, 
\sqrt{1+\frac{\partial_\psi\eta^2}{\eta^2} +\left(\mathcal{H}[\psi,\eta]\right)^2\frac{\partial_x\eta^2 }{\eta^4}}\ , 
\\
\mathcal{H}[\psi,\eta]&=&\, 2 \bz \cosh\left[ \frac{1}{4} \log\left(1+\frac{\eta^2}{\bz^4}
+\frac{2\eta}{\bz^2}\cos\psi \right)\right]\ , \rule{0pt}{.8cm}\\
\mathcal{K}[\psi,\eta]&=& \frac{1}{\eta}\left( \sqrt{1+\bz^4\eta^2+2\bz^2\eta\cos\psi}\right)\ . 
\label{K}
\rule{0pt}{.8cm}
\eea
The reader can find the Lagrangian for the D3 case in Appendix \ref{APPEQ}. Details about 
the equation of motion are not important, and numerical studies of the solution will be 
presented elsewhere. In this section, we focus mainly on the role of the separatrix, 
and discuss how to describe (globally) the \RT surface. 

The starting point is similar to that of Section \ref{PERTURBSECTION}. We know that the 
equation of motion of $\eta(\rad,\psi)$ depends on the dimensionful parameter $\bz$, and 
we want to exploit the scale invariance of the IR fixed point by writing a suitable ansatz 
for the solution. The idea is to recover the IR solution (\ref{IRESOLUNO}) in the limit 
$\bz\to\infty$, therefore we consider\footnote{Notice that this is a different ansatz compared to 
the UV \eqref{UVANSATZ}.}
\beq\label{IRANSATZ}
\eta(\rad,\psi)=  \frac{ \bz }{\ell} \Func(\yc,\psi)
~.
\eeq  
The equation of motion for the field $\Func(\yc,\psi)$ depends on a single dimensionless 
parameter $\frac{\Delta}{R^2}=\frac{\bz \ell}{R^2}$. The limit $\frac{\Delta}{R^2}\to\infty$ is 
well defined and gives back the equation (\ref{IREQUATIONZERO}). It is then possible to 
set up a perturbative calculation in inverse powers of $\Delta$ whose form is
\beq\label{FUNCIREXPANSION}
\Func(\yc,\psi)=\frac{2 R^2_{IR}}{\sqrt{1-\yc^2}}+\sum_{k=1}^{\infty} \frac{1}{\Delta^k} f^{(k)}_p(\yc,\theta)
\, .
\eeq
The functions $f^{(k)}$ would be determined at each order in perturbation theory. However, 
unlike the UV expansion, now the perturbative series breaks down in some range of $\yc$.
We can understand this point in two ways. One way is to realize that the expansion in inverse 
powers of $\Delta$ that we are using involves, for example, expressions like
\beq
{\sqrt{1+\frac{\Func^2(\yc,\psi)}{\Delta^2} +\frac{2\Func(\yc,\psi)}{\Delta}\cos\psi}} 
= 1+ \sum_i c_i(\psi) \left(\frac{\Func(\yc,\psi)}{\Delta}\right)^i
\eeq
(see e.g.\ \eqref{K}). Therefore, it would be strictly valid as long as $\Func(\yc,\psi)<\Delta$ for any 
$\yc, \psi$. Problems arise with this requirement when $\yc\to 1$ because the surface is 
approaching the UV boundary and $\Func$ diverges. 

The second argument relies on the observation that the functions $f^{(k)}$ will generically 
diverge faster than $\eta_{UV}\sim 1/{(1-\yc^2)}$, thus violating the known UV $AdS$ 
asymptotics. For example, in our D1-D5 system the first $f^{(k)}$ that we find 
are\footnote{In writing these $f^{(k)}$ we are imposing one boundary condition, 
$\partial_\yc f^{(k)}=0$ at $\yc=0$, and we are fixing the remaining integration constant 
to some value. In principle we should keep this integration constant and use it as a 
matching parameter. However, the argument we want to make here does not 
depend on this choice.}
\bea
\Func(\yc,\psi)&=&\frac{2 R^2_{IR}}{\sqrt{1-\yc^2}}
+ \frac{v}{1-\yc^2} \left(\frac{R^2_{IR}}{\Delta}\right) +
\nn\\
&&+\frac{3}{4}\frac{v^2}{\left(1-\yc^2\right)^{3/2}} 
\left(\frac{R^2_{IR}}{\Delta}\right)^2+
\frac{v \left(v^2-3(1-\yc^2)\right)}{\left(1-\yc^2\right)^2}\left(\frac{R^2_{IR}}{\Delta}\right)^3\, 
+\ldots 
\eea
where $v=\cos\psi$. The second line involves powers higher than $(1-\yc^2)^{-1}$.

From these observations we conclude that the perturbative expansion 
(\ref{FUNCIREXPANSION}) is a good approximation of the solution only below a certain 
$\yc_s$, potentially related to the existence of the separatrix. The right way to recover the 
UV solution is to make use of a matched expansion.

Before discussing the matching procedure at the UV boundary, we would like to make the 
following comment. In the limit $\Delta\to\infty$, it is clear that the separatrix becomes a 
UV cut-off and the full geometry breaks into the sum of two disconnected throats. Such 
fragmentation is nicely understood in the $(u,v)$ plane as the process of zipping the upper 
from the lower half plane (the dashed line on the left plot of Figure \ref{FIGURE3} on the 
$u$-axis moves off to infinity). 
However, for $\Delta\gg1$ but finite, the IR geometry is still connected all the way 
up to the UV and the separatrix is the natural short distance cut-off from the deep 
IR perspective. The resummation of the series \eqref{CORRECTIR} seems to be in direct
relation with the resummation of an infinite set of irrelevant interactions that one has to 
perform in the effective IR field theory to reconstruct the whole RG flow.

The matching expansion is based on the assumption that as we zoom into the boundary 
region $\yc\to 1$ we effectively look into the UV $AdS$. In order to do so, it is standard 
to define both a rescaled variable $\bar{\yc}=(1-\yc)/\epsilon$ and a rescaled function 
$\Func=\epsilon^{\alpha}\bar{\Func}$, and take the limit $\epsilon\to0$ in the equation 
of motion. In such a limit, the new variable $\bar{\yc}$ and new function $\bar{\Func}$ 
are kept fixed. Because $\Func$ diverges at the boundary $\alpha$ has to be negative. 
In our case we know that $\alpha=-2$ because we are taking a limit in which the theory 
is conformal and we know the scalings. As a result, the matching procedure gives back 
$\eta_{UV}$ with an overall constant that we need to determine. By inspection of the 
equation of motion we find that
\beq
\Func=\frac{1}{\Delta} \frac{R^4_{UV}}{1-\yc^2}\ .
\eeq 
The matched expansion leads to an expression of the form
\beq\label{RESULTMATCHIR1}
\Func(\yc,\psi)= \frac{2 R^2_{IR}}{\sqrt{1-\yc^2}}
+ \frac{1}{\Delta}\,  \frac{R^4_{UV}}{1-\yc^2} 
+ \big[ \ldots\mathrm{\, matched \ expansion \ corrections}\ldots \big]
~.
\eeq
Returning to the original embedding field $\eta(\yc,\psi)$ we find
\beq\label{RESULTMATCHIR2}
\eta(\yc,\psi)= \frac{\bz}{\ell} \frac{2 R^2_{IR}}{\sqrt{1-\yc^2}} 
+\frac{1}{\ell^2} \frac{R^4_{UV}}{1-\yc^2} 
+\big[ \ldots\mathrm{\, matched \ expansion \ corrections}\ldots \big]\ .
~,
\eeq

\subsection{Entanglement fragmentation}
Inserting the solution \eqref{RESULTMATCHIR1} into the entropy functional we can calculate 
the leading large-$\ell$ behavior of the holographic entanglement entropy in the Coulomb 
branch RG flow. The resulting expression will give the correct expectation: 
the HEE receives one contribution from the UV $AdS_{p+2}$ (with radius $R_{UV}$), 
and another one from the two disconnect IR $AdS_{p+2}$ (with radius $R_{IR}$). 
In the following, we will make this statement more precise by splitting the integration over 
$\yc\in[0,\ell)$ into an IR and a UV contribution. 

It is useful to define the integral
\beq
I_p[\yh_{min},\yh_{max}]=\int_{\yh_{min}}^{\yh_{max}}ds\ \frac{\left(1-\yh^2\right)^{(p-2)/2}}{\yh^p}\ .
\eeq
We already encountered $I_p$ in Section \ref{HEE2uv}. 
In particular, $I_p[\frac{a}{\ell},1]$ calculates the $\ell$ dependence of the HEE of spherical 
entangling surfaces for pure $AdS_{p+2}$.

In the limit $\Delta\to\infty$, the form of the solution \eqref{RESULTMATCHIR1} implies that the 
HEE is that of two $AdS_{p+2}$ with radius $R_{IR}$, as expected
\beq\label{FACTOR2}
\frac{S_p}{\mathcal{A}_p}\,({\ell\to\infty})\, =\, 2\, C^{IR}_p\, R_{IR}^{p}\,  I_p\left[\frac{a}{\ell},1\right]
\eeq
where $\mathcal{A}_p$ is the area of the entangling surface and
\beq
C^{IR}_p=\frac{1}{4 G_N^D } \, \mathrm{Vol}(S^{D-p-2})\, R_{IR}^{D-p-2} .
\eeq
The factor of $2$ in \eqref{FACTOR2} counts the two disconnected $AdS$ throats, and comes 
from the integration over the angle $\psi$. For a generic multi-centered configuration with 
$K$ IR throats the result will be given in terms of the sum of $K$ contributions. The 
integration over $\yh$ needs the UV regulator $a/\ell$, as usual in $AdS$. Notice that 
this cut-off is the one that regulates the volume of the IR $AdS$ after taking the decoupling limit.

At $\Delta\gg 1$ the exact solution of $\eta(\yc,\psi)$ will exhibit a separatrix and thus we 
need to consider the matched expansion. We can estimate roughly that the IR solution 
becomes sub-leading compared to the UV at 
\beq\label{CUTOFF}
X_c\,\approx\,\frac{R_{UV}^4}{2R^2_{IR}}\frac{1}{\Delta}\,
=\,\frac{R_{UV}^4}{2 \bar{z}\, R^2_{IR}}\frac{1}{\ell}\,\equiv\,\frac{\bar{a}}{\ell}\ ,
\eeq
where $X=1-\yc^2$. Therefore it is useful to separate the integration over $\yc$ in a UV 
contribution, in which we can use $\eta\approx\eta_{UV}$, and an IR contribution, in which 
we can use $\eta\approx\eta_{IR}$. In our approximation, this way of splitting the integral 
over $\yc$ isolates the contributions coming from below and above the separatrix. This 
is a natural thing to do because in the limit $\Delta\to\infty$ the separatrix will become the UV 
cut-off. The final result for the HEE is
\bea\label{IREEAPPROX}
\frac{S_p}{\mathcal{A}_p} &=&  2\, C^{IR}_p\, R_{IR}^{p}\, I_p\left[\frac{\bar{a}}{\ell},1\right]\, +\, 
C^{UV}_p\, R_{UV}^{p}\, I_p\left[\frac{a}{\ell},\frac{\bar{a}}{\ell}\right]\, +\, \ldots
\eea
where $a$ is a UV cut-off, $\bar{a}$ can be read from \eqref{CUTOFF}, and finally
\beq
C^{UV}_p=\frac{1}{4 G_N^D } \, \mathrm{Vol}(S^{D-p-2})\, R_{UV}^{D-p-2}\ .
\eeq
The result \eqref{IREEAPPROX} agrees with the general expectations for the HEE along RG flows 
\cite{Liu:2013una,Albash:2011nq}.

\section{More about the connectivity index in the IR effective theory}
\label{moreIR}

The behavior of the entanglement entropy that we studied in previous sections suggests that the
change of the connectivity index along the RG flow is a process with sharp features at intermediate energies in the large-$N$ limit. 
The discussion in sections \ref{intro} and \ref{fragment} suggests that the origins of these features can be traced 
to the qualitatively different properties of the theory at large and small energies. In particular, we pointed out that part of the 
interaction between the IR CFTs at small energies is mediated by multi-trace operators,
and stated that such interactions cannot change the IR connectivity index in the large-$N$ limit.
Any change of the IR connectivity index must be driven by the interactions with the singleton degrees
of freedom. Since this is one of the main points of the proposed picture we would like to summarize here some 
well known facts that support its validity.

\paragraph{Energy-momentum conservation, bi-gravity and the connectivity index.}
Let us consider the flow from $SU(N)$ to $SU(N_1)\times SU(N_2) \times U(1)$ in the large-$N$
limit. As described in section \ref{fragment}, in quantum field theory the infrared effective description of this flow
involves two large-$N$ IR CFTs deformed separately by single-trace interactions denoted schematically by $V_I$ in \eqref{multiaa}. 
Interaction between these theories comes from multi-trace interactions of the schematic form $\OO_1 \OO_2$ in \eqref{multiaa}
and from interactions with the abelian singleton degrees of freedom. In this section we want to examine what would happen 
in the large-$N$ limit if the interaction with the singleton degrees of freedom were absent.
All the interactions are IR-irrelevant, which means that one has to work with an explicit UV cutoff
both in field theory and the AdS/CFT correspondence.

Refs.\ \cite{Kiritsis:2006hy,Aharony:2006hz} demonstrated that multi-trace interactions alone do not introduce any anomalous dimensions to the 
two energy-momentum tensors of the deformed IR product theory at leading order in $1/N$ in the large-$N$ limit.\footnote{In 
\cite{Kiritsis:2006hy,Aharony:2006hz} this statement was shown for double-trace deformations involving
scalar single-trace operators, but it is not hard to show in general that the leading correction to the
anomalous dimension of the energy-momentum tensors is $1/N$ suppressed as a consequence of the large-$N$ counting.}
As a result, even after the deformation, the theory continues to have two separately conserved energy-momentum
tensors. This is the first sign that the connectivity index cannot be modified as we increase the
energy if the singletons do not contribute in the IR effective field theory description. Notice that the subleading $1/N$ corrections 
introduce an anomalous dimension to a linear combination of the energy-momentum tensors
and the connectivity index necessarily gets reduced. 

Ignoring the contribution of singletons, on the holographically dual side the IR effective description involves a bi-gravity (bi-string) theory 
\cite{Kiritsis:2006hy,Aharony:2006hz,Kiritsis:2008at} with the following features. The spacetime of
{\it each} graviton asymptotes towards the UV to a deformed $AdS \times S$ space. The UV deformation
introduces the $`1'$ in the harmonic function of each throat as we expand the full harmonic function of
the double-center solution around each center. This deformation captures the irrelevant 
single-trace part of the deformations $V_I$ in each theory mentioned previously.\footnote{From the 
UV point of view the IR bi-gravity description arises as we localize the wavefunction of the single graviton
in the multi-center geometry in the vicinity of each center.}
In addition, at leading order in $1/N$, the multi-trace deformations impose modified boundary
conditions for the fields in the bulk \cite{Witten:2001ua,Berkooz:2002ug}. It was shown in 
\cite{Kiritsis:2006hy,Aharony:2006hz} that the bulk gravitons remain massless at leading order
and the bi-gravity theory is trivial (namely, besides the modified boundary conditions, the
theory in the bulk is a decoupled product of string theories living on separate spacetimes with 
separate Lagrangians). Subleading $1/N$ corrections make a linear combination of the bulk
gravitons massive (i.e.\ modify the gravity Lagrangians) and reduce the connectivity index. 

The above discussion suggests that the effects of the multi-trace interactions alone
do not reduce the IR connectivity index at leading order in $1/N$. The effects that are responsible
for this reduction at the planar level come from the interaction with the abelian singleton degrees 
of freedom. In the bulk bi-gravity picture these are interactions that take place on the boundary 
and make the sources dynamical. In the presence of these interactions only one combination 
of the bulk stress-energy momentum tensors is classically conserved.

\paragraph{Factorizability of correlation functions.} We mentioned in the introduction that one of the
signs of separability is factorization in correlation functions. Here we would like to examine how separability 
and factorization of correlation functions work at leading order in $1/N$ in a large-$N$ product theory deformed 
only by multi-trace deformations. For example, the presence of the double-trace inter-CFT interactions 
in \eqref{multiaa} modifies the correlation functions already at leading order in the $1/N$ expansion. 
In particular, the correlation function $\langle \OO_1(x_1) \OO_2(x_2) \rangle$ (recall $\OO_i$, $i=1,2$, 
is a single-trace operator in the IR CFT$_i$) receives $h_{12}$ contributions and is no longer vanishing. 
This effect alone seems to spoil the extreme IR factorizability, so one may wonder how
it is consistent with the above-proposed separability in the vicinity of the IR fixed point in the absence
of singleton contributions.

It is perhaps simpler to describe the resolution of this question in AdS/CFT language along the following
lines. For concreteness, let us focus on two single-trace (scalar) operators $\OO_1$, $\OO_2$
and assume for clarity that the total effective field theory action is
\beq
\label{moreIRaa}
\SS_{total} = \SS_1 + \SS_2 + \int d^{p+1}x \, h_{12} \OO_1 \OO_2
~.
\eeq
$\SS_1$, $\SS_2$ are the actions of two CFTs, CFT$_1$ and CFT$_2$.
In the bulk bi-gravity theory there are two scalar fields, $\phi_1$ and $\phi_2$, corresponding
to $\OO_1$, $\OO_2$. With the boundary of each $AdS$ spacetime at large radius $r_i$ $(i=1,2)$, each of these fields
will asymptote to 
\beq
\label{moreIRab}
\phi_i = \frac{\alpha_i}{r_i^{\Delta_i}} +\ldots 
+ \frac{\beta_i}{r_i^{p+1-\Delta_i}}+\ldots
~.
\eeq 
$\Delta_i$ is the scaling dimension of the operator $\OO_i$.
Assuming $\Delta_i > \frac{p+1}{2}$ the double-trace deformation on the r.h.s.\ of equation \eqref{moreIRaa}
is irrelevant. Also, with this assumption the $\beta_i$ term in \eqref{moreIRab} is the leading term as 
$r_i \to \infty$.
 
The generating functional of the theory \eqref{moreIRaa} is obtained by adding sources for $\OO_i$,
\beq
\label{moreIRac}
\delta \SS = \int d^{p+1} x \, \left( J_1 \OO_1 + J_2 \OO_2 \right)
~,
\eeq
and computing the quantum path integral of the full theory 
\beq
\label{moreIRad}
Z = e^{-W[J_1,J_2]}
\eeq
as a function of the sources $J_i$.
Then, connected correlation functions of $\OO_1$, $\OO_2$ are computed by functional derivatives
of $W$ with respect to $J_i$.

In gravity one computes the on-shell gravity action $I_{GR}$ as a function of the asymptotic
coefficients $\beta_i$ in \eqref{moreIRab}. In the case at hand, these obey the boundary conditions
\beq
\label{moreIRae}
\beta_1 = J_1 + h_{12}\, \alpha_2~, ~~
\beta_2 = J_2 + h_{12}\, \alpha_1
~.
\eeq
Using the conditions coming from the regularity of the bulk solutions within the framework
of designer (bi)gravity \cite{Hertog:2004ns,Kiritsis:2011zq} one can fix a second pair of relations 
between $\beta_1$ and $\alpha_1$ on the one hand, and $\beta_2$ and $\alpha_2$ on the other. This allows 
to re-express the bulk solution and the corresponding on-shell gravity action in terms 
of $J_i$. Since the bulk theory is a direct product of two gravity theories
\beq
\label{moreIRaf}
I_{GR} [J_1,J_2]= I_{GR,1}[J_1,J_2] + I_{GR,2}[J_1,J_2]
~.
\eeq

The basic relation of the AdS/CFT correspondence is
\beq
\label{moreIRag}
W[J_1,J_2] = I_{GR}[J_1,J_2]
~. 
\eeq
Because of \eqref{moreIRaf}, \eqref{moreIRag} we see, for example, that the correlation functions 
$\langle \OO_1(x_1) \OO_2(x_2)\rangle$ are non-vanishing and factorizability is seemingly lost.
However, the above procedure reveals that the main effect of the double-trace deformation is to 
mix the sources $J_i$. Denoting the new combinations as $\tilde J_i \equiv \beta_i$, so that 
\beq
\label{moreIRai}
W[\tilde J_1, \tilde J_2] = I_{GR,1}[\tilde J_1] + I_{GR,2}[\tilde J_2]
~,
\eeq
we see that there is a new basis of operators (dual to $\tilde J_i$) where factorization of correlation 
functions reappears. The new basis is non-trivially related to the old one with redefinitions
of the form
\beq
\label{moreIRaia}
\OO_1 = \frac{\delta \tilde J_1}{\delta J_1} \tilde \OO_1 + \frac{\delta \tilde J_2}{\delta J_1} \tilde \OO_2 
~, ~~
\OO_2 = \frac{\delta \tilde J_1}{\delta J_2} \tilde \OO_1 + \frac{\delta \tilde J_2}{\delta J_2} \tilde \OO_2
\eeq
at any $J_1, J_2$. For correlation functions we need to take at the end $J_1,J_2\to 0$. When the 
regularity conditions are linear, e.g.\ $\beta_i = f_i\, \alpha_i$ for constant $f_i$, 
the coefficients of the linear transformation \eqref{moreIRaia} are simple functions of the parameters 
$f_1,f_2, h_{12}$. For non-linear regularity conditions, e.g.\ $\beta_i = f_i\, \alpha_i^{p_i}$ with $p_i$ positive
real exponents, the same coefficients are algebraically less straightforward to obtain. We have computed
them for $p_1=p_2=2$ as functions of $f_1,f_2, h_{12}$, but the expressions are not particularly illuminating
and will not be presented here explicitly.

Although these arguments do not examine the correlation functions of the most general operators,
combined with the statements about energy-momentum conservation, they motivate the
expectation that it is possible to find density matrices that obey the relation
\beq
\label{moreIRaj}
\rho_A = \rho_{A,1} \otimes \rho_{A,2}
~,
\eeq
by defining appropriately the Hilbert spaces $H_{A,1}$ and $H_{A,2}$ (over which we trace) 
in order to account for the new basis of operators identified in \eqref{moreIRaia}. 
Equivalently, we expect that the corresponding relative quantum entropy continues to vanish
in the deformed theory \eqref{moreIRaa}, $S_{12}(\rho_A) =0$, and that the connectivity index remains 2.

\section{Discussion}
\label{discussion}

Generic processes rearrange the interactions and correlations between different degrees of 
freedom in a quantum system. In some cases the Hilbert space experiences a fragmentation 
where the interaction between degrees of freedom in different parts of the system becomes 
weak or even disappears.\footnote{The inverse is also possible. The interactions between 
different parts of a fragmented Hilbert space may turn on and grow.} When the latter happens 
correlation functions appropriately factorize and we say that the process 
changed the connectivity index of the system.

In this paper we pointed out that there are instances where such processes 
can exhibit sharp features at finite interaction coupling.
We examined a particular class of examples that occur in the Coulomb branch of large-$N$ 
superconformal field theories. In that class we presented evidence that 
suggests that the effect is a consequence of a competition between
large-$N$ effects and effects associated to the specifics of the renormalization group flow.
It would be interesting to learn if there are other classes of quantum systems that exhibit this 
kind of behavior. A potentially interesting holographic context for this purpose 
is the context of Ref.\ \cite{Lin:2004nb}.

We discussed two major probes of transitions between
fixed points with different connectivity indices. The first one is entanglement entropy on a spatial
region $A$ and the second one are quantum information measures of separability, e.g.\ relative
entropy of entanglement and quantum mutual information. The main lessons and emerging open
questions of our study can be summarized as follows.

\paragraph{Entanglement entropy.}
For spherical regions the entanglement entropy $S$ is a function of the radius $\ell$ of the 
sphere. We computed this function in the Coulomb branch of large-$N$ gauge theories and 
noticed that a sharp feature appears through the formation of a separatrix in the 
Ryu-Takayanagi surface. The separatrix is absent for $\ell < \ell_c$ and present for any 
$\ell \geq \ell_c$, where $\ell_c$ is a critical radius. The presence of $\ell_c$ signals a change 
in the behavior of the entanglement above $\ell_c$, but since we lack an analytic solution of 
the Ryu-Takayanagi surface in all regimes, it has been hard to determine the precise nature 
of this change. It would be very interesting to learn if the entanglement entropy is a $C^\infty$ 
function at $\ell_c$, or whether some derivative of $S$ diverges.

It would also be important to understand better why the perturbative UV results \eqref{EEUVD3}, 
\eqref{EEUVD1D5}, and \eqref{EEUVM2}, do not depend on $\Delta^2$ and $\Delta^3$ corrections to the RT surface. 
In field theory, it is natural to associate those contributions to operators of dimension 
$2$ and $3$. The perturbative holographic computation would be reliable for small entangling 
regions and one way to proceed would be to develop a small length OPE expansion for the twist 
fields. Because of supersymmetry some coefficients in the OPE may be directly vanishing, or 
may vanish when the limit $n\to 1$ in the replica trick is taken. 
This would also provide a non-trivial check of the RT prescription out of conformality.    

\paragraph{Entanglement measures of separability.} We pointed out that the quantum information 
notion of separability is a very suitable probe of physics in processes that change the connectivity 
index. In our examples we expect the quantum mutual information $S_{12}(\rho_{A})$
to vanish at $\ell=\infty$ (the extreme IR) and increase as $\ell$ decreases. We also expect
certain suppressing effects in the large-$N$ limit.

It would be interesting to know:
\begin{itemize}

\item[$(a)$] if $S_{12}(\ell)$ exhibits a critical radius $\ell^*_c$, analogous
to $\ell_c$ of the entanglement entropy, and if so, what is
the precise relation between the critical radii $\ell_c$ and $\ell^*_c$, e.g.\ whether 
$\ell_c = \ell^*_c$. Also, we would like to determine the precise behavior of $S_{12}(\ell)$ at $\ell^*_c$, 
e.g.\ in order to verify whether it is continuous at that point, or whether some derivative diverges. 

\item[$(b)$] it would be useful for many general purposes to know how to compute 
$S_{12}(\ell)$ efficiently, for instance with holographic methods in the AdS/CFT correspondence. 
Notice that the definition of $S_{12}(\ell)$ involves the entanglement entropy $S(\ell)$, that can be 
computed holographically \`a la Ryu-Takayanagi, and the entanglement entropies of the 
reduced density matrices $\rho_{A,1}$, $\rho_{A,2}$. The authors of the recent paper 
\cite{Mollabashi:2014qfa} argued that the latter entropies for $A^c =  {\O}$ are computed 
in $AdS \times S$ spacetimes by a co-dimension-2 surface that goes through the equator 
of the transverse sphere $S$. It would be interesting to know if there is a generalization 
of this statement for $A^c \neq {\O}$. Related questions and quantities have been 
discussed in the recent condensed matter literature in \cite{Furukawa,Xu,Chen,Lundgren}.

\end{itemize}

Similar observations and questions can be made for other measures of separability, for example
the relative entropy of entanglement $D_{\rm REE}$ \eqref{entad},
although most likely this is a much harder quantity to compute explicitly.

\section*{Acknowledgments}

\noindent We would like to thank V.\ Cardoso, O.\ Dias, L.\ Di Pietro, S.\ Hartnoll, T.\ Ishii, E.\ Kiritsis, H.\ Okawa, G.\ Policastro, 
J.\ Russo, T.\ Tufarelli. The work of F.A. and V.N. was supported in part by European Union's Seventh 
Framework Programme under grant agreements (FP7-REGPOT-2012-2013-1) no 316165, the EU-Greece
program ``Thales" MIS 375734 and was also co-financed by the European
Union (European Social Fund, ESF) and Greek national funds through the Operational
Program ``Education and Lifelong Learning" of the National Strategic Reference
Framework (NSRF) under ``Funding of proposals that have received a positive
evaluation in the 3rd and 4th Call of ERC Grant Schemes".

\appendix

\section{Minimal surface equations} 
\label{APPEQ}

In all cases analyzed in the main text, the Lagrangian of the \RT minimal surface
can be put into the form
\beq
\mathcal{L}\,=\, \rad^{p-1}\, \mathcal{K}[\theta,\eta]\, \mathcal{H}[\theta,\eta]\, 
\sqrt{1+ \frac{1}{\alpha}\frac{\partial_{\theta}\eta^2}{ \eta^2} 
+ \left(\mathcal{H}[\theta,\eta]\right)^2\, \partial_\rad \eta^2}
\eeq
where $\alpha$ is a constant. We will use the notation $\partial_\rad \eta=\eta^{(1,0)}$ 
and $\partial_\theta \eta=\eta^{(0,1)}$. For example, the case of D3 branes in the 
coordinates of section \ref{HEE2ir} is
\bea
\mathcal{H}^2&=& \frac{1}{\eta^4} \frac{1}{4\sqrt{\bz^4+\eta^2+2\bz^2\eta\cos\psi}} 
\left[ 2\left( \bz^2+\sqrt{\bz^4+\eta^2+2\bz^2\eta\cos\psi}\right)^2-\eta^2\right]
~,
\\
\mathcal{K}&=& \eta^3 \left[ \sqrt{1+\frac{\bz^4}{\eta^2} +\frac{2\bz^2}{\eta}\cos\psi } 
+\left(\frac{\bz^2}{\eta}+\cos\psi\right)\right]^2 
~.
\rule{0pt}{.8cm}
\eea

The equation of motion is quite complicated and can be expressed as the sum of 
different pieces. We found convenient to write it as
\beq
D_0+D_1+D_2+D_3=0
~.
\eeq
The first operator, $D_0$, is a generalization of the flat space minimal surface equation, namely
\beq
D_0=\, -d_{(2,0)} \eta^{(2,0)} - d_{(1,1)} \eta^{(1,1)}-d_{(0,2)}\eta^{(0,2)}
+\frac{1}{\alpha\eta} \left[ \frac{\eta^{(0,1)} }{\eta\, \FF}\right]^2 
\eeq
with
\beq
d_{(2,0)}=1+\left[ \frac{\eta^{(0,1)} }{\eta} \right]^2~,\qquad
d_{(1,1)}= -2\,\frac{\eta^{(0,1)}\eta^{(1,0)} }{\eta^2}~,\qquad
d_{(0,2)}= \frac{1}{\eta^2\, \FF^2} +\left[\frac{\eta^{(1,0)}}{\eta}\right]^2
~.
\eeq
The remaining terms are
\beq
D_1 = \frac{1}{\FF^2} \left[ 1+ \frac{\left(\eta^{(0,1)}\right)^2}{\alpha\eta^2} 
+ \left( \eta^{(1,0)} \right)^2\FF^2 \right]
\left( \frac{\KK^{(0,1)}}{\KK} - \frac{\eta^{(0,1)}}{\alpha\eta^2} \frac{\KK^{(1,0)}}{\KK} \right)
~,
\eeq
\beq
D_2 = -\frac{1}{2}\left(1+\frac{\left( \eta^{(0,1)} \right)^2}{\alpha\eta^2}\right)
\left[ \partial_\eta\left(\frac{1}{\FF^2}\right) 
- \frac{\eta^{(0,1)}}{\alpha\eta^2}\partial_\theta\left(\frac{1}{\FF^2}\right) \right]
~,
\eeq
\beq
D_3=\frac{1-d}{x}\eta^{(1,0)} \left[ 1+ \frac{\left(\eta^{(0,1)}\right)^2}{\alpha\eta^2} 
+ \left( \eta^{(1,0)} \right)^2\FF^2 \right]
~.
\eeq

\addtocontents{toc}{\protect\setcounter{tocdepth}{1}}
\addtocontents{lof}{\protect\setcounter{tocdepth}{1}}

\bibliography{EE} 

\providecommand{\href}[2]{#2}\begingroup\raggedright\begin{thebibliography}{10}

\bibitem{Holzhey:1994we}
C.~Holzhey, F.~Larsen, and F.~Wilczek, {\it {Geometric and renormalized entropy
  in conformal field theory}},  {\em Nucl.Phys.} {\bf B424} (1994) 443--467,
  [\href{http://xxx.lanl.gov/abs/hep-th/9403108}{{\tt hep-th/9403108}}].

\bibitem{Calabrese:2009qy}
P.~Calabrese and J.~Cardy, {\it {Entanglement entropy and conformal field
  theory}},  {\em J.Phys.} {\bf A42} (2009) 504005,
  [\href{http://xxx.lanl.gov/abs/0905.4013}{{\tt arXiv:0905.4013}}].

\bibitem{Myers:2010tj}
R.~C. Myers and A.~Sinha, {\it {Holographic c-theorems in arbitrary
  dimensions}},  {\em JHEP} {\bf 1101} (2011) 125,
  [\href{http://xxx.lanl.gov/abs/1011.5819}{{\tt arXiv:1011.5819}}].

\bibitem{Aharony:2006hz}
O.~Aharony, A.~B. Clark, and A.~Karch, {\it {The CFT/AdS correspondence,
  massive gravitons and a connectivity index conjecture}},  {\em Phys.Rev.}
  {\bf D74} (2006) 086006, [\href{http://xxx.lanl.gov/abs/hep-th/0608089}{{\tt
  hep-th/0608089}}].

\bibitem{Kiritsis:2006hy}
E.~Kiritsis, {\it {Product CFTs, gravitational cloning, massive gravitons and
  the space of gravitational duals}},  {\em JHEP} {\bf 0611} (2006) 049,
  [\href{http://xxx.lanl.gov/abs/hep-th/0608088}{{\tt hep-th/0608088}}].

\bibitem{Intriligator:1999ai}
K.~A. Intriligator, {\it {Maximally supersymmetric RG flows and AdS duality}},
  {\em Nucl.Phys.} {\bf B580} (2000) 99--120,
  [\href{http://xxx.lanl.gov/abs/hep-th/9909082}{{\tt hep-th/9909082}}].

\bibitem{Costa:1999sk}
M.~S. Costa, {\it {Absorption by double centered D3-branes and the Coulomb
  branch of N=4 SYM theory}},  {\em JHEP} {\bf 0005} (2000) 041,
  [\href{http://xxx.lanl.gov/abs/hep-th/9912073}{{\tt hep-th/9912073}}].

\bibitem{Ryu:2006bv}
S.~Ryu and T.~Takayanagi, {\it {Holographic derivation of entanglement entropy
  from AdS/CFT}},  {\em Phys.Rev.Lett.} {\bf 96} (2006) 181602,
  [\href{http://xxx.lanl.gov/abs/hep-th/0603001}{{\tt hep-th/0603001}}].

\bibitem{Nishioka:2006gr}
T.~Nishioka and T.~Takayanagi, {\it {AdS Bubbles, Entropy and Closed String
  Tachyons}},  {\em JHEP} {\bf 0701} (2007) 090,
  [\href{http://xxx.lanl.gov/abs/hep-th/0611035}{{\tt hep-th/0611035}}].

\bibitem{Klebanov:2007ws}
I.~R. Klebanov, D.~Kutasov, and A.~Murugan, {\it {Entanglement as a probe of
  confinement}},  {\em Nucl.Phys.} {\bf B796} (2008) 274--293,
  [\href{http://xxx.lanl.gov/abs/0709.2140}{{\tt arXiv:0709.2140}}].

\bibitem{wilde}
M.~M. Wilde, {\it {Quantum Information Theory}},  {\em Cambridge University
  Press} (2013) 669 pgs, [\href{http://xxx.lanl.gov/abs/1106.1445}{{\tt
  arXiv:1106.1445}}].

\bibitem{Hartnoll:2014ppa}
S.~A. Hartnoll and R.~Mahajan, {\it {Holographic mutual information and
  distinguishability of Wilson loop and defect operators}},
  \href{http://xxx.lanl.gov/abs/1407.8191}{{\tt arXiv:1407.8191}}.

\bibitem{Mollabashi:2014qfa}
A.~Mollabashi, N.~Shiba, and T.~Takayanagi, {\it {Entanglement between Two
  Interacting CFTs and Generalized Holographic Entanglement Entropy}},  {\em
  JHEP} {\bf 1404} (2014) 185, [\href{http://xxx.lanl.gov/abs/1403.1393}{{\tt
  arXiv:1403.1393}}].

\bibitem{Morita:2011cs}
T.~Morita and V.~Niarchos, {\it {F-theorem, duality and SUSY breaking in
  one-adjoint Chern-Simons-Matter theories}},  {\em Nucl.Phys.} {\bf B858}
  (2012) 84--116, [\href{http://xxx.lanl.gov/abs/1108.4963}{{\tt
  arXiv:1108.4963}}].

\bibitem{Kiritsis:2008at}
E.~Kiritsis and V.~Niarchos, {\it {Interacting String Multi-verses and
  Holographic Instabilities of Massive Gravity}},  {\em Nucl.Phys.} {\bf B812}
  (2009) 488--524, [\href{http://xxx.lanl.gov/abs/0808.3410}{{\tt
  arXiv:0808.3410}}].

\bibitem{Aharony:2008ug}
O.~Aharony, O.~Bergman, D.~L. Jafferis, and J.~Maldacena, {\it {N=6
  superconformal Chern-Simons-matter theories, M2-branes and their gravity
  duals}},  {\em JHEP} {\bf 0810} (2008) 091,
  [\href{http://xxx.lanl.gov/abs/0806.1218}{{\tt arXiv:0806.1218}}].

\bibitem{Maldacena:1997re}
J.~M. Maldacena, {\it {The Large N limit of superconformal field theories and
  supergravity}},  {\em Int.J.Theor.Phys.} {\bf 38} (1999) 1113--1133,
  [\href{http://xxx.lanl.gov/abs/hep-th/9711200}{{\tt hep-th/9711200}}].

\bibitem{Kraus:1998hv}
P.~Kraus, F.~Larsen, and S.~P. Trivedi, {\it {The Coulomb branch of gauge
  theory from rotating branes}},  {\em JHEP} {\bf 9903} (1999) 003,
  [\href{http://xxx.lanl.gov/abs/hep-th/9811120}{{\tt hep-th/9811120}}].

\bibitem{Klebanov:1999tb}
I.~R. Klebanov and E.~Witten, {\it {AdS / CFT correspondence and symmetry
  breaking}},  {\em Nucl.Phys.} {\bf B556} (1999) 89--114,
  [\href{http://xxx.lanl.gov/abs/hep-th/9905104}{{\tt hep-th/9905104}}].

\bibitem{Skenderis:2006uy}
K.~Skenderis and M.~Taylor, {\it {Kaluza-Klein holography}},  {\em JHEP} {\bf
  0605} (2006) 057, [\href{http://xxx.lanl.gov/abs/hep-th/0603016}{{\tt
  hep-th/0603016}}].

\bibitem{Costa:2000gk}
M.~S. Costa, {\it {A Test of the AdS / CFT duality on the Coulomb branch}},
  {\em Phys.Lett.} {\bf B482} (2000) 287--292,
  [\href{http://xxx.lanl.gov/abs/hep-th/0003289}{{\tt hep-th/0003289}}].

\bibitem{Dimopoulos:2001ui}
S.~Dimopoulos, S.~Kachru, N.~Kaloper, A.~E. Lawrence, and E.~Silverstein, {\it
  {Small numbers from tunneling between brane throats}},  {\em Phys.Rev.} {\bf
  D64} (2001) 121702, [\href{http://xxx.lanl.gov/abs/hep-th/0104239}{{\tt
  hep-th/0104239}}].

\bibitem{Dimopoulos:2001qd}
S.~Dimopoulos, S.~Kachru, N.~Kaloper, A.~E. Lawrence, and E.~Silverstein, {\it
  {Generating small numbers by tunneling in multithroat compactifications}},
  {\em Int.J.Mod.Phys.} {\bf A19} (2004) 2657--2704,
  [\href{http://xxx.lanl.gov/abs/hep-th/0106128}{{\tt hep-th/0106128}}].

\bibitem{Casini:2011kv}
H.~Casini, M.~Huerta, and R.~C. Myers, {\it {Towards a derivation of
  holographic entanglement entropy}},  {\em JHEP} {\bf 1105} (2011) 036,
  [\href{http://xxx.lanl.gov/abs/1102.0440}{{\tt arXiv:1102.0440}}].

\bibitem{Lewkowycz:2013nqa}
A.~Lewkowycz and J.~Maldacena, {\it {Generalized gravitational entropy}},  {\em
  JHEP} {\bf 1308} (2013) 090, [\href{http://xxx.lanl.gov/abs/1304.4926}{{\tt
  arXiv:1304.4926}}].

\bibitem{Witten:1998zw}
E.~Witten, {\it {Anti-de Sitter space, thermal phase transition, and
  confinement in gauge theories}},  {\em Adv.Theor.Math.Phys.} {\bf 2} (1998)
  505--532, [\href{http://xxx.lanl.gov/abs/hep-th/9803131}{{\tt
  hep-th/9803131}}].

\bibitem{Klebanov:2000hb}
I.~R. Klebanov and M.~J. Strassler, {\it {Supergravity and a confining gauge
  theory: Duality cascades and chi SB resolution of naked singularities}},
  {\em JHEP} {\bf 0008} (2000) 052,
  [\href{http://xxx.lanl.gov/abs/hep-th/0007191}{{\tt hep-th/0007191}}].

\bibitem{Minahan:1998xq}
J.~Minahan and N.~Warner, {\it {Quark potentials in the Higgs phase of large N
  supersymmetric Yang-Mills theories}},  {\em JHEP} {\bf 9806} (1998) 005,
  [\href{http://xxx.lanl.gov/abs/hep-th/9805104}{{\tt hep-th/9805104}}].

\bibitem{Lang}
S.~Lang, {\it {Complex analysis}},  {\em Springer-Verlag} {\bf 4th edition}
  (1999).

\bibitem{Liu:2013una}
H.~Liu and M.~Mezei, {\it {Probing renormalization group flows using
  entanglement entropy}},  {\em JHEP} {\bf 1401} (2014) 098,
  [\href{http://xxx.lanl.gov/abs/1309.6935}{{\tt arXiv:1309.6935}}].

\bibitem{Albash:2011nq}
T.~Albash and C.~V. Johnson, {\it {Holographic Entanglement Entropy and
  Renormalization Group Flow}},  {\em JHEP} {\bf 1202} (2012) 095,
  [\href{http://xxx.lanl.gov/abs/1110.1074}{{\tt arXiv:1110.1074}}].

\bibitem{Witten:2001ua}
E.~Witten, {\it {Multitrace operators, boundary conditions, and AdS / CFT
  correspondence}},  \href{http://xxx.lanl.gov/abs/hep-th/0112258}{{\tt
  hep-th/0112258}}.

\bibitem{Berkooz:2002ug}
M.~Berkooz, A.~Sever, and A.~Shomer, {\it {'Double trace' deformations,
  boundary conditions and space-time singularities}},  {\em JHEP} {\bf 0205}
  (2002) 034, [\href{http://xxx.lanl.gov/abs/hep-th/0112264}{{\tt
  hep-th/0112264}}].

\bibitem{Hertog:2004ns}
T.~Hertog and G.~T. Horowitz, {\it {Designer gravity and field theory effective
  potentials}},  {\em Phys.Rev.Lett.} {\bf 94} (2005) 221301,
  [\href{http://xxx.lanl.gov/abs/hep-th/0412169}{{\tt hep-th/0412169}}].

\bibitem{Kiritsis:2011zq}
E.~Kiritsis and V.~Niarchos, {\it {Josephson Junctions and AdS/CFT Networks}},
  {\em JHEP} {\bf 1107} (2011) 112,
  [\href{http://xxx.lanl.gov/abs/1105.6100}{{\tt arXiv:1105.6100}}].

\bibitem{Lin:2004nb}
H.~Lin, O.~Lunin, and J.~M. Maldacena, {\it {Bubbling AdS space and 1/2 BPS
  geometries}},  {\em JHEP} {\bf 0410} (2004) 025,
  [\href{http://xxx.lanl.gov/abs/hep-th/0409174}{{\tt hep-th/0409174}}].

\bibitem{Furukawa}
S.~Furukawa and Y.~B. Kim, {\it {Entanglement entropy between two coupled
  Tomonaga-Luttinger liquids}},  {\em Phys.Rev.} {\bf B83} (2011) 085112,
  [\href{http://xxx.lanl.gov/abs/1009.3016}{{\tt arXiv:1009.3016}}].

\bibitem{Xu}
C.~Xu, {\it {Entanglement Entropy of Coupled Conformal Field Theories and Fermi
  Liquids}},  {\em Phys.Rev.} {\bf B84} (2011) 125119,
  [\href{http://xxx.lanl.gov/abs/1102.5345}{{\tt arXiv:1102.5345}}].

\bibitem{Chen}
X.~Chen and F.~E., {\it {Quantum Entanglement and Thermal Reduced Density
  Matrices in Fermion and Spin Systems on Ladders}},  {\em J.Stat.Mech} (2013)
  P08013, [\href{http://xxx.lanl.gov/abs/1305.6538}{{\tt arXiv:1305.6538}}].

\bibitem{Lundgren}
R.~Lundgren, Y.~Fuji, S.~Furukawa, and M.~Oshikawa, {\it {Entanglement spectra
  between coupled Tomonaga-Luttinger liquids: Applications to ladder systems
  and topological phases,}},  {\em Phys.Rev} {\bf B88} (2013) 245137,
  [\href{http://xxx.lanl.gov/abs/1310.0829}{{\tt arXiv:1310.0829}}].

\end{thebibliography}\endgroup
  
\end{document}